\documentclass[twocolumn,showpacs,preprintnumbers, 
amsmath,amssymb,prb,superscriptaddress]{revtex4}

\usepackage[T1]{fontenc}
\usepackage[latin1]{inputenc}
\usepackage{graphicx}
\usepackage{color}
\usepackage{hyperref}

\newcommand{\etal}{\mbox{\textit{et al.}}}

\newcommand{\cf}{{cf.~}}

\newcommand{\Eqref}[1]{Eq.~(\ref{#1})}
\newcommand{\Figref}[1]{Fig.~\ref{#1}}
\newcommand{\Secref}[1]{Sec.~\ref{#1}}

\begin{document}
\title{On the identification of pristine and defected graphene nanoribbons \\
by phonon signatures in the electron transport characteristics}

\author{Rasmus B. Christensen}
\affiliation{Dept. of Micro- and Nanotechnology, Technical University of Denmark, {\O}rsteds Plads, Bldg.~345E, DK-2800 Kongens
Lyngby, Denmark}

\author{Thomas~Frederiksen}
\affiliation{Donostia International Physics Center (DIPC) -- UPV/EHU, Donostia-San Sebasti\'an, Spain}
\affiliation{IKERBASQUE, Basque Foundation for Science, Bilbao, Spain}

\author{Mads~Brandbyge}
\affiliation{Dept. of Micro- and Nanotechnology, Technical
University of Denmark, {\O}rsteds Plads, Bldg.~345E, DK-2800 Kongens
Lyngby, Denmark}
\email{mads.brandbyge@nanotech.dtu.dk}
\date{\today}

\begin{abstract}
Inspired by recent experiments where electron transport was measured across graphene nanoribbons (GNR) 
suspended between a metal surface and the tip of a scanning tunneling microscope
[Koch \textit{et al}.,~Nat.~Nanotechnol.~{\bf 7}, 713 (2012)], we present detailed first-principles 
simulations of inelastic electron tunneling spectroscopy (IETS) of long pristine and defected 
armchair and zigzag nanoribbons under a range of charge carrier conditions.
For the armchair ribbons we find two robust IETS signals around 169 and 196 mV corresponding to the D- and G-modes
of Raman spectroscopy as well as additional fingerprints due to various types of defects in the edge passivation. 
For the zigzag ribbons we show that the spin state strongly influences the spectrum and thus propose IETS
as an indirect proof of spin polarization.
\end{abstract}
\maketitle

\begin{figure*}  
\includegraphics[trim= 0cm 0cm 0cm 0cm,clip=true,width=1.7\columnwidth]{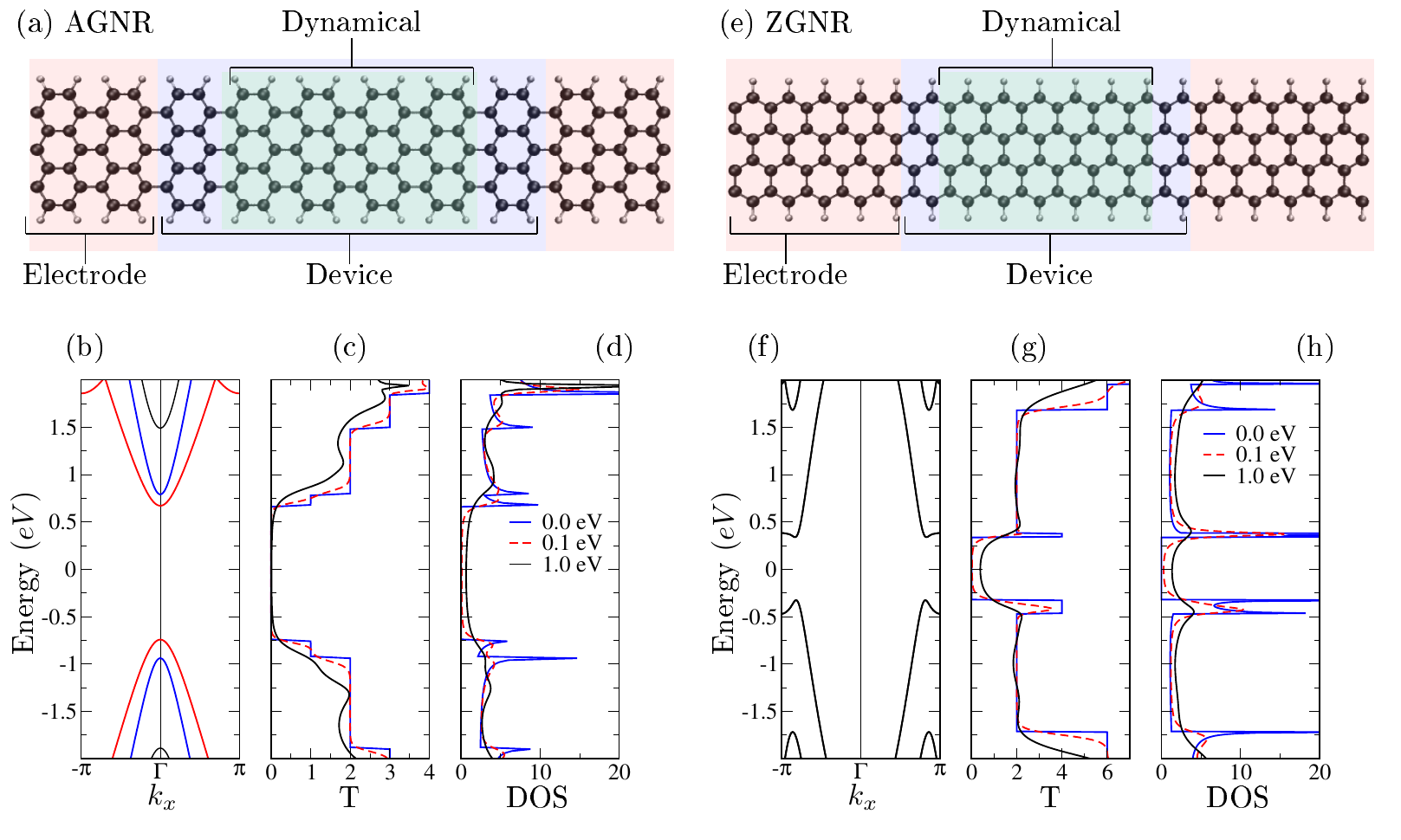}\vspace{-0.5cm}
\caption{(Color online) (a) Computational setup for a pristine AGNR showing electrode, device and dynamical regions.  (b) Electronic band structure ($k_x$ is in units of inverse unit cell length). The different bands are colored according to symmetry of the electronic states. Red: symmetric, corresponding to \Figref{fig:channels}(a-b). Blue: anti-symmetric, corresponding to \Figref{fig:channels}(c-d). (c) Electronic transmission for varying electrode broadening describing the coupling to the metal contacts, $\eta=0,0.1,1$ eV, see text. (d) Electronic DOS projected onto the dynamical region. Panels (e)-(h) show the similar entities for the pristine ZGNR case.}
\label{fig:CleanGNR}
\end{figure*}

\begin{figure}  
\includegraphics[trim= 0cm 0cm 0cm 0cm,clip=true,width=\columnwidth]{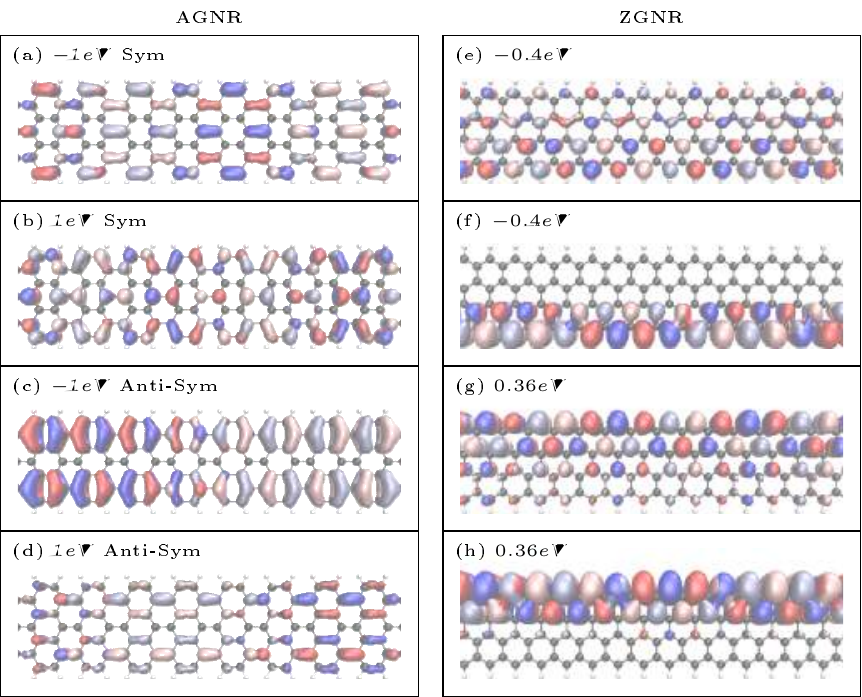}
\caption{(Color online) (a)-(d) Electron transmission eigenchannels for the clean AGNR for the valence bands at $E-E_F=-1$ eV and for the conduction bands at $E-E_F=1$ eV. (e)-(h) Electron transmission eigenchannels for the clean ZGNR in the valence bands at $E-E_F=-0.4$ eV and in the conduction bands at $E-E_F=0.4$ eV for one spin component. The eigenchannels for the other spin component are simply mirror images around the middle of the ZGNR (not shown). The red/blue (pink/gray) isosurfaces represent the real (imaginary) part and sign of the scattering state wave function.
For all eigenchannel calculations the electrode broadening was set to zero ($\eta=0$ eV). 
}
\label{fig:channels}
\end{figure}

\section{\label{sec:Intro}Introduction}

Graphene as the basis of a new generation of electronics\cite{Raza2011,FoaTorres2014} has been the center of much attention in the last years, and devices based on 
nanostructured graphene have been put forward. The most generic form of nanostructured graphene is graphene nanoribbons (GNR),\cite{Dutta2010} and other structures, such as graphene anti-dot lattices\cite{pedersen_graphene_2008,Bai2010}, can be viewed as networks of them. GNRs are potential candidates for molecular wires with tailored conductance properties. For graphene-based nanostructures the edges and their passivation, as well as defects inside the structure, can play crucial roles for the transport properties.\cite{Wagner2013}
However, characterization of edge passivation or structural/chemical defects is challenging especially after device fabrication. Raman spectroscopy\cite{Ferrari2013} can give information about defects on large areas of the sample, while tip-enhanced Raman spectroscopy (TERS)\cite{Shiotari2014} have been used in combination with STM on GNRs. However, Raman studies involve averages over larger areas (> 10 nm), 
and does not yield information about the impact of vibrations on transport. In that aspect inelastic electron tunneling spectroscopy (IETS) serves as a way of performing non-destructive characterization yielding vibrational/phonon fingerprints of a range of defects. In order to interpret IETS experiments, theoretical modeling of the inelastic signals in the electronic current due to electron-phonon (e-ph) scattering is needed.

GNRs have been fabricated using different strategies including lithographic techniques,\cite{Han2007} chemical synthesis,\cite{Li2008a,Wu2010} epitaxial growth\cite{Baringhaus2014}, and longitudinal unzipping of carbon nanotubes.\cite{Kosynkin2009} Furthermore, several groups have succeeded in atomically precise bottom-up fabrication of armchair GNRs (AGNR)\cite{Cai2010,Blankenburg2012}, chiral GNRs,\cite{Han2014} and AGNR hetero-junctions\cite{Cai2014} grown on metal surfaces. Experimentally, the vibrational properties have been investigated by Raman spectroscopy and the electronic structure has been mapped out by STM, angle-resolved (two-photon) photo-emission and high-resolution electron energy loss spectroscopy.\cite{Bronner2012,Ruffieux2012,Shiotari2014} Signatures of phonon excitation were observed by STM in the differential conductance spectroscopy performed at the zigzag termini state of AGNRs adsorbed on Au(111), and these signatures were shown to be sensitive to modifications in the local atomic geometry.\cite{VanderLit2013f} 
AGNRs have also been lifted up from the weakly bonding Au(111) surface with the tip of a STM enabling measurements of the voltage-dependent conductance in suspended configurations.\cite{Koch2012}  

From the theoretical side density-functional theory (DFT) has been used to investigate the stability of structural and chemical reconstructions of GNR edges,\cite{Wassmann2008,Li2010c,Wagner2013a} together with the transport and band-gap engineering.\cite{Hod2007,Gunlycke2007,Zhang2012a,Al-Aqtash2013,Wagner2013} The vibrational properties and phonon band structure have been calculated with empirical potentials\cite{Yamamoto2004} and DFT.\cite{Vandescuren2008,Gillen2009} In addition, there have been theoretical predictions\cite{Zhou2007,Saito2010a} of the Raman spectrum, in good agreement with experiments.\cite{Huang2012b,Cai2010} For a finite AGNR the role of zigzag termini states have been studied theoretically, comparing DFT to the many-body Hubbard model.\cite{Ijas2013}

Inspired by the recent lifting experiments by Koch \etal,\cite{Koch2012} we here investigate theoretically the signals of e-ph scattering in the conductance of long GNRs between metal electrodes. Our aim is two-fold. First, we want to address the role phonon scattering in the transport characteristics of pristine GNRs. Second, we wish to compute detailed IETS for different GNRs under varying charge carrier concentrations 
and explore how different types of realistic defects may modify the IETS and thus possibly be directly probed in transport measurements. We focus on the two most generic edge types, namely armchair (AGNR) and zigzag (ZGNR), and pay attention to the effects of spin polarization in the latter case. In actual experiments the substrate or an applied gate potential control the Fermi level $E_F$ in the ribbons. To address this variability we scan $E_F$ using a numerically effective scheme enabling fast calculations of the IETS.\cite{Lu2014}
We find that the AGNR generally display two robust IETS signals around $169$ and $196$ mV corresponding to the D- and G-modes of Raman spectroscopy and that a dehydrogenated dimer at the edge should further leave a clear defect signal at around $245$ mV.
For the ZGNR we find that the spin polarization breaks the mirror symmetry around the middle of the ribbon resulting in IETS signals from a range of modes around the D- and G-mode energies. For both AGNR and ZNGR defects which break the planar symmetry of ribbons allows for contributions to the IETS from out-of-plane phonon modes.

The paper is organized as follows. First we discuss our atomistic model setup for the density functional and electron transport calculations, and outline the approach for the IETS simulations. In Sec.~III we present our results for pristine AGNR and ZGNR and relate their transport properties and IETS to the band structures. In Sec.~IV we turn to the defected systems by considering realistic possibilities of defects in the edge passivation, backbone bonding motifs, and presence of adatoms. Finally, a summary and our conclusions are presented in Sec.~V.

\section{\label{sec:Method}Methods}

We calculate the electronic and vibrational structure from DFT
using the academic codes \textsc{Siesta}/\textsc{TranSiesta}.\cite{Soler2002,Brandbyge2002}
We employ the generalized gradient approximation (GGA) for the exchange-correlation functional,\cite{PeBuEr.96}
a single-zeta polarized (SZP) basis set for the carbon and hydrogen atoms, and use a cut-off energy of 400-500 Ry for the real-space grid. 
These choices, balancing accuracy and computational cost, provide a
good description to investigate trends and general behavior of the substantial number of systems
considered in this work.

The vibrational degrees of freedom, calculated by diagonalization of the dynamical matrix, and the e-ph couplings are extracted from finite differences as implemented in the \textsc{Inelastica} code.\cite{Paulsson2005,Frederiksen2007,Inelastica}
The armchair and zigzag GNRs considered here are shown in \Figref{fig:CleanGNR}. 
We adopt the usual two-probe setup with the device region ($D$) coupled to left ($L$) and right ($R$) electrodes with all electronic matrix elements expressed in a local basis set.
The primitive unit cell of the AGNR (ZGNR) consists of 18 (10) atoms and in our calculations this unit cell is repeated 10 (18) times in the transport direction to form the scattering regions illustrated in \Figref{fig:CleanGNR}(a,e). 
The electrode couplings $\mathbf{\Gamma}_{L/R}$ 
are included on the two first/last unit cells before folding onto $D$. 
In our treatment a subset of atoms in $D$ is allowed to vibrate. We fix this dynamical region, restricted by the condition that the e-ph couplings are fully included inside $D$, to the 4 and 6 central unit-cells for the AGNR and ZGNR, respectively. 
The corresponding e-ph couplings used to calculate the inelastic electron transport are thus expressed in the center 6 unit-cells for the AGNR and 8 unit-cells for the ZGNR. The convergence of our results with the size of the dynamical region is addressed below.

We generally consider nanoribbons that are suspended between two metallic leads. In the case of the lifting experiments,\cite{Koch2012} these would correspond to the metal sample surface and the STM tip. We wish here to focus on the action inside the GNRs and put aside the possible complications due to the detailed electronic structure of the metals, and the metal-GNR interface in particular. To this end we introduce a simple model of the metal electrodes without substantial electronic features: we use semi-infinite GNRs with highly broadened states (effectively smearing out energy gaps). In practice this is done by adding a finite numerical imaginary part $\eta$ to the energy argument in the electrode recursion calculation.\cite{Sancho1984}
This scheme ensures that the phonon effects originate from the GNRs themselves and not from details of the metal-GNR interface, which is generally unknown in the STM experiments. The electronic band structures for the infinite ribbons, along with the transmission and density of states (DOS) are shown for $\eta=0,0.1,1$~eV in \Figref{fig:CleanGNR}(b,c,d) and  Fig.~\ref{fig:CleanGNR}(f,g,h) for AGNR and ZGNR, respectively. We note that the broadened transmission spectrum [\Figref{fig:CleanGNR}(d)] is quite consistent with the experimentally reported differential conductance curves for AGNR.\cite{Koch2012} 
The electronic states involved in the transport are shown in \Figref{fig:channels} in terms of the transmission eigenchannels\cite{Paulsson2007} in the valence and conduction bands of the AGNR and ZGNR. Their spatial symmetry play a significant role for the selection rules involved in the inelastic scattering as discussed later.

In principle, the electronic structure should be evaluated at finite bias. However, without a detailed model of the connection to the metal electrodes (where an important part of the voltage drop will take place) and for sufficiently long systems (in which the electric field will be small), it is reasonable to use the zero-voltage electronic structure 
and to simply assume a symmetric voltage drop over the two identical, idealized device-electrode interfaces.
More specifically, in the following we consider that the chemical potentials of the electrodes move according to the applied bias voltage $V$ and possibly an applied gate voltage $V_G$ (mimicking actual doping or electrostatic gating that modify the charge carrier concentration in an experimental setup) according to $\mu_{L/R}=E_F\pm{\hbar\omega_{\lambda}}/{2}+V_G$.

\subsection{Computational scheme for IETS}
For a device strongly coupled to the electrodes, a coupling between the electron current $I(V)$ and a phonon mode $\lambda$ ideally shows up at zero temperature as a step discontinuity in the differential conductance when the inelastic phonon emission process becomes energetically allowed, that is, when the chemical potential difference exceeds the quantum of vibrational energy, $|\mu_L-\mu_R|=\hbar\omega_\lambda$. Thus, around the emission threshold the electronic states involved in the scattering process are those at $\mu_L$ and $\mu_R$. The IETS signal, conventionally expressed as the ratio between the second and first derivatives of the current with respect to the voltage,
\begin{equation}
\mathrm{IETS} = \frac{\partial_{V}^2{I}(V)}{\partial_{V}{I}(V)},
\end{equation}
is calculated by considering the e-ph coupling as the perturbation on the current, evaluated using the nonequilibrium Green's functions (NEGF). In the so-called lowest order expansion (LOE) the
inelastic part of the differential conductance can be written as,\cite{Lu2014}
\begin{align}
\label{eq:current}
\partial_{V}{I}(V)=&  \gamma_\lambda \, \partial_{V}{\cal I}^\mathrm{sym}(V,\hbar\omega_\lambda, T)
\\
 &+  \kappa_\lambda\,\partial_{V}{\cal I}^\mathrm{asym}(V,\hbar\omega_\lambda, T),\nonumber
\end{align}
where summation over the vibration index $\lambda$ is assumed. ${\cal I}^\mathrm{sym}$ and ${\cal I}^\mathrm{asym}$ are the ``universal'' (system-independent) functions that depend on the applied bias $V$, phonon energy $\hbar\omega_\lambda$ and the temperature $T$.
Assuming the electronic and phononic distribution functions are given by the Fermi-Dirac and Bose-Einstein distributions, respectively, their analytical expressions can be written as:
\begin{align}
	\mathcal{I}^\mathrm{sym}\! \equiv&\! \frac{\mathrm{G}_0}{2e}\sum_{s=\pm} \!\!\!s(\hbar\omega_\lambda+s eV) \label{eq:sym}
	\\
&\times\!\left(\!\coth\!\frac{\hbar\omega_ \lambda}{2k_BT}-\coth\!\frac{\hbar\omega_\lambda+s eV}{2k_BT}\!\right)\nonumber,
\\
	\mathcal{I}^\mathrm{asym} \equiv& \frac{\mathrm{G}_0}{2e}\int_{-\infty}^{+\infty}\!d\varepsilon\mathcal{H}\{ f(\varepsilon'_-)- f(\varepsilon'_+)\}(\varepsilon)	\label{eq:asym}\\
	&\times\left[ f(\varepsilon-eV)-f_{}(\varepsilon)\right],\nonumber
\end{align}
where $\mathrm{G}_0=2e^2/h$ is the conductance quantum,  $f(\varepsilon)$ is the Fermi-Dirac function, $\varepsilon'_s\equiv\varepsilon'+s\hbar\omega_\lambda$, and $\mathcal H$ denotes the Hilbert transform.

The signal amplitudes $\gamma_\lambda$ and $\kappa_\lambda$ of the symmetric and antisymmetric signals in the differential conductance are even and odd in bias, respectively. For a symmetric structure the asymmetric signal vanishes in the wide-band approximation (LOE-WBA)\cite{Paulsson2005}. However, this is not guaranteed in the more general treatment employed here,\cite{Lu2014} where the energy dependence of the electronic structure is explicitly taken into account. The amplitudes $\gamma_\lambda$ and $\kappa_\lambda$ are expressed in terms of electronic structure quantities and e-ph couplings,\cite{Lu2014}
\begin{align}
\gamma_{\lambda}=&{\rm Tr}\![ \mathbf M_\lambda \tilde{\mathbf A}_L(\mu_L) \mathbf M_\lambda \mathbf A_R(\mu_R) ]+{\rm Im} B_\lambda, \label{eq:gamma}
\\
\kappa_\lambda =& 2{\rm Re} B_\lambda, \label{eq:kappa}
\end{align}
where $B_\lambda $ is defined as
\begin{align}
B_\lambda \equiv &{\rm Tr}[\mathbf M_\lambda\mathbf A_R(\mu_L)\mathbf\Gamma_L(\mu_L)\mathbf G^r(\mu_L)\mathbf M_\lambda\mathbf A_R(\mu_R) \nonumber
\\
&-\mathbf M_\lambda \mathbf G^a(\mu_R)\mathbf\Gamma_L(\mu_R)\mathbf A_R(\mu_R)\mathbf M_\lambda \mathbf A_L(\mu_L)]. \label{eq:ic4ag}
\end{align}
In the above, ${\bf M}_\lambda$ denotes the e-ph coupling matrix for mode $\lambda$, $\mathbf G^{r/a}$ the retarded/advanced unperturbed Green's functions, and  ${\bf A}_\alpha=\mathbf G^{r}\mathbf\Gamma_\alpha\mathbf G^{a}$ the spectral density matrices for left/right moving states with the time-reversed version $\tilde{\mathbf A}_\alpha=\mathbf G^{a}\mathbf\Gamma_\alpha\mathbf G^{r}$.
The purely electronic quantities are thus being evaluated at the chemical potentials of the left/right electrodes corresponding to the excitation threshold for each vibration.
We compute ${\bf M}_\lambda$ with the finite-difference scheme of \textsc{Inelastica} taking the vacuum energy as a common reference (in absence of real metal leads to pin the Fermi energy).\cite{Frederiksen2007}

In the localized atomic basis set of \textsc{Siesta} all the above quantities are matrices defined in the electronic space corresponding to region $D$. The second derivatives of the universal functions in Eqs.~(\ref{eq:sym})-(\ref{eq:asym}) are sharply peaked around the phonon threshold. For this reason the coefficients $\gamma_{\lambda}$ and $\kappa_\lambda$ can be considered voltage-independent with their values computed exactly at the threshold. Due to the computational efficiency of the LOE scheme described above we are able to evaluate the IETS on a fine grid of gate voltages $V_G$ spanning a large range of relevant values between valence and conduction bands of the GNRs.

\section{Pristine graphene nanoribbons}\label{sec:Results}

Now we first turn to the IETS results of the two pristine (clean) ribbons, and in the following section to the impact of selected defects in the IETS. As our main system we focus on the AGNR systems directly relevant for the lifting experiments.\cite{Koch2012} The results for the ZGNR are provided mainly as comparison and to look into the role of chirality and in particular effects rooted in spin polarization, and thus we now discuss these separately.

\begin{figure} 
\includegraphics[trim= 0cm 0cm 0cm 0cm,clip=true,width=1.\columnwidth]{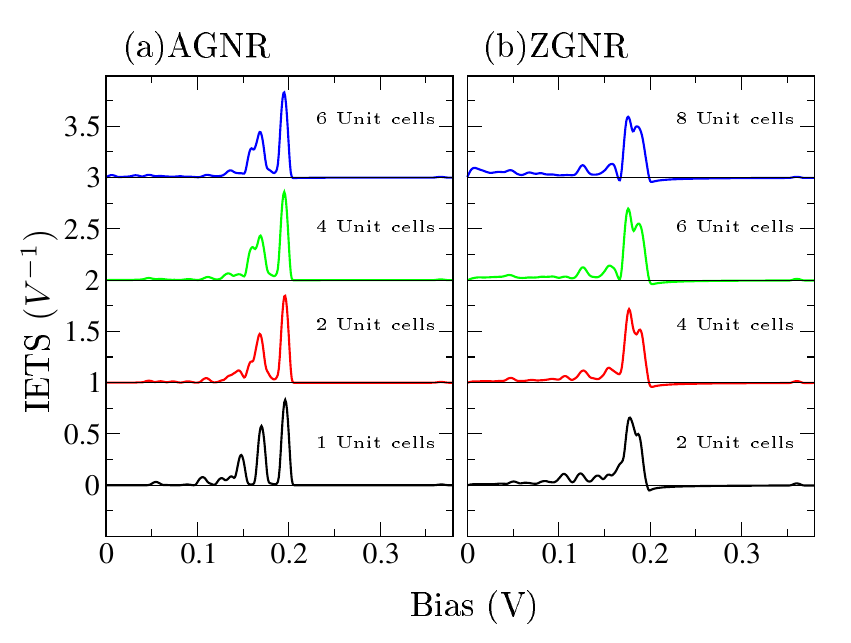}\vspace{-0.3cm}
\caption{\label{fig:ConvClean} (Color online)  Convergence of the intrinsic IETS for pristine (a) AGNR and (b) ZGNR
as a function of the size of the dynamical region (stated in the legends). The results are normalized with respect to the number of vibrating unit cells, i.e., we show the IETS amplitude per H$_4$C$_{14}$ segment for AGNR and per H$_2$C$_8$ segment for ZGNR. No gate voltage is applied ($V_G=0.0$ V).}
\end{figure}

\begin{figure} 
	\includegraphics[trim= 0cm 0cm 0cm 0cm,clip=true,width=1.\columnwidth]{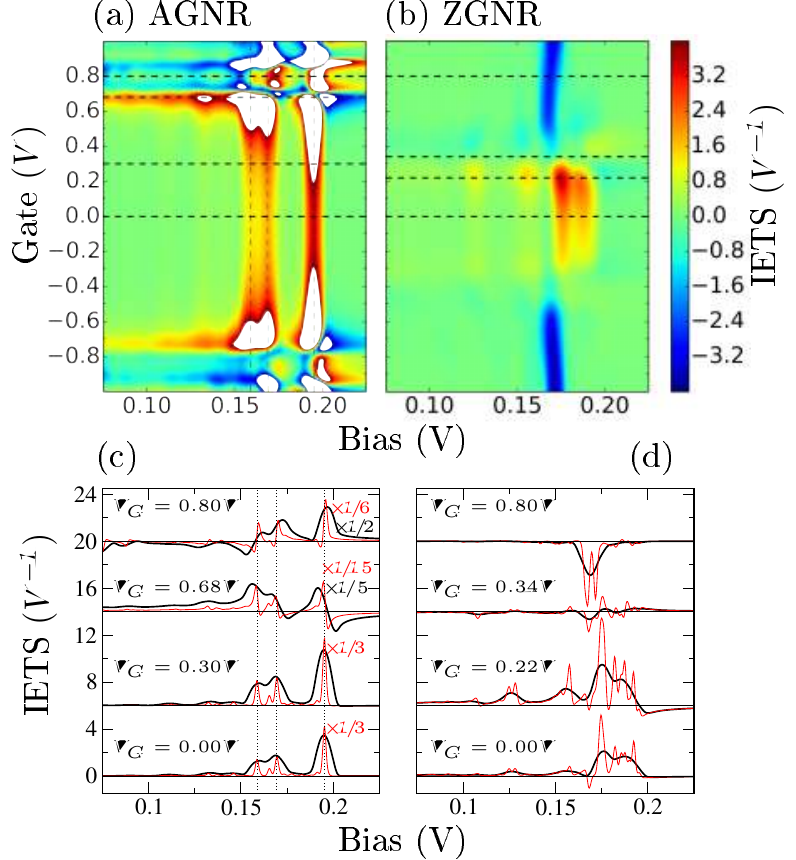}
	\caption{\label{fig:IETSclean}(Color online) IETS signals as a function of gate voltage for (a) pristine AGNR (4 vibrating unit cells) and (b) pristine ZGNR (6 vibrating unit cells).
	Vertical dashed lines are guides to the eye indicating the energy of the most contributing vibrational modes. Specific IETS signals for the (c) AGNR and (d) ZGNR at selected gate voltages
    marked with horizontal dashed lines in panels (a) and (b). 
    Broadening originates from temperature $T=4.2$ K and a lock-in modulation voltage $V_\mathrm{rms}=5$ mV
    (except for the thin red lines in the lower panels with $V_\mathrm{rms}=0$~mV ).}
\end{figure}

\begin{figure} 
\includegraphics[trim= 0cm 0cm 0cm 0cm,clip=true,width=\columnwidth]{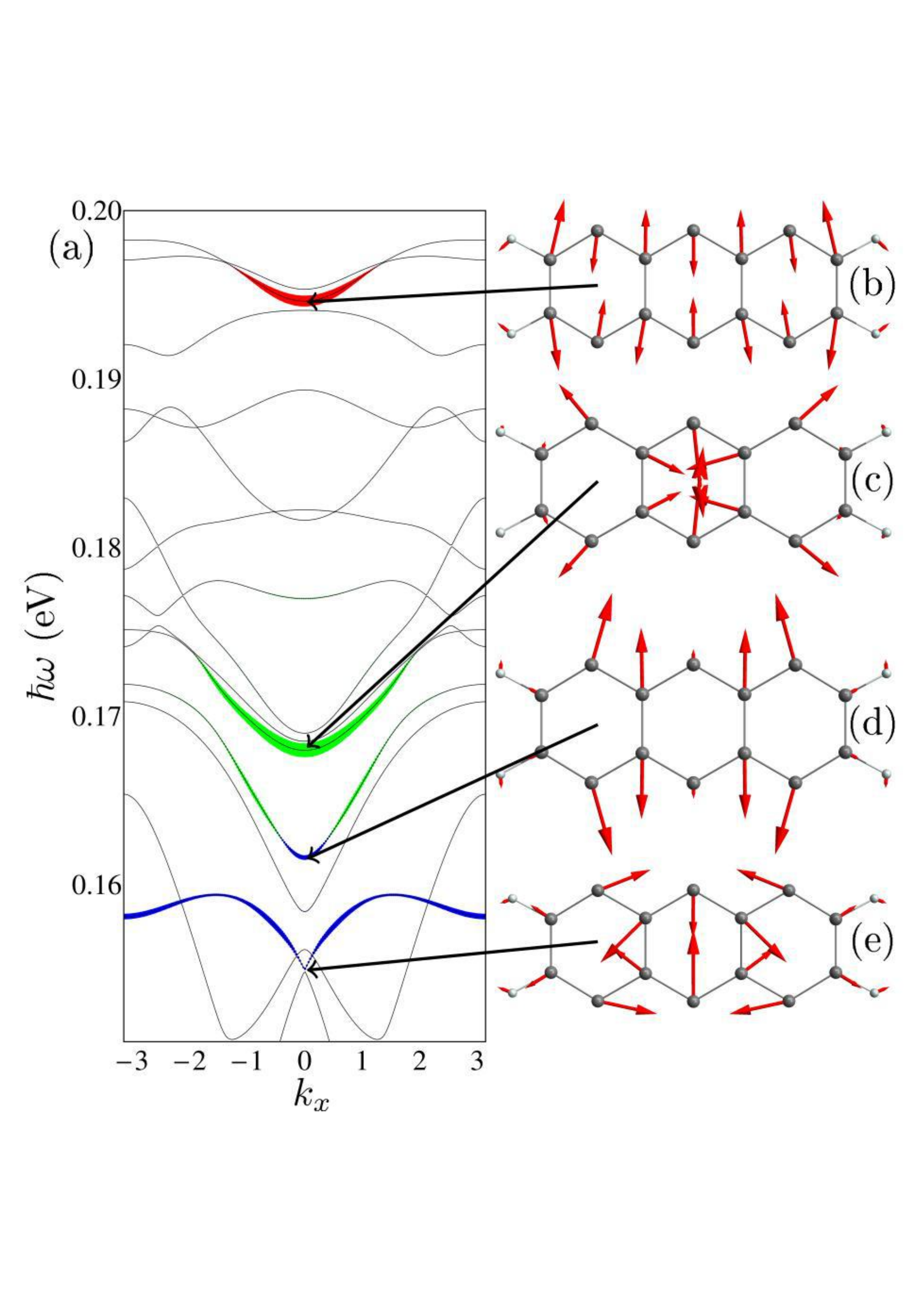}
\caption{\label{fig:AGNRphononband} (Color online) (a) Computed phonon band structure for the pristine, infinite AGNR  ($k_x$ is in units of inverse unit cell length). 
The magnitude of the red, green and blue bands (corresponding to the three vertical lines in \Figref{fig:IETSclean}(c)), is proportional to the signal size weighted overlap, ($F_{nk}(V_G=0\mathrm{V})$ in \Eqref{eq.weightF}), between the repeated band vector and modes with frequencies $\hbar\omega>180$ meV, $180>\hbar\omega>162$ meV and $\hbar\omega<162$ meV for red, green and blue, respectively. The red band is scaled by $0.2$ compared to blue and green. 
(b-e) Selected phonon band modes at $\Gamma$ for the infinite structure which, according to the projection, characterize the active IETS modes. Units of $k_x$?}
\end{figure}

\subsection{Pristine armchair nanoribbons} \label{subsec:CleanAGNR} 
As representative of the AGNR class we have investigated a pristine AGNR with a width of $W=7$ dimers (7-AGNR) corresponding to a C-C edge distance of 7.5 {\AA} (see \Figref{fig:CleanGNR}). It presents a direct semi-conducting band gap $E_g$ due to the lateral confinement and can be classified as a ``large-gap ribbons'' since $p=2$ is an integer in the relation $W=3p+1$.\cite{Raza2011} We obtain $E_g \approx 1.3$~eV at the present level of approximation (DFT-GGA and SZP basis set), as seen from the electronic band structure shown in \Figref{fig:CleanGNR}(b). This value is smaller than 
those estimated experimentally ($E_g \approx$ 2.3-2.6~eV for a flat AGNR on Au(111) \cite{Ruffieux2012,Bronner2014} and $E_g\approx 2.7$ eV for an AGNR suspended between surface and STM-tip\cite{Koch2012}) due to the underestimation of electron-electron interaction\cite{Yang2007} which plays an more important role in quasi one-dimensional GNRs compared to pristine graphene. Dielectric screening from the substrate also influences significantly the actual gap size:
a band gap of $3.2$~eV for a 7-AGNR was found to be lowered to $2.7$~eV on a hexagonal boron-nitride (hBN) substrate using GW calculations,\cite{Jiang2013} similar to the lowering calculated for a 7-AGNR on Au(111).\cite{Liang2012} In general we expect that underestimation of band gaps would mainly amount to a simple scaling the Fermi level position within the gap.

We first discuss the effect of the finite size of the dynamical region in our treatment. Figure \ref{fig:ConvClean}(a)
shows how the IETS signals for the AGNR (at fixed gate voltage $V_G=0.0$ V) vary as a function of the size of the dynamical region, ranging from 1 to 6 unit cells. For easy comparison, the data are normalized by the number of vibrating unit cells. As the signal amplitudes in this representation are roughly constant we conclude that the absolute IETS simply scale linearly with the active e-ph coupling region. Consequently, the magnitudes in IETS may thus provide insight into the active scattering region in actual experiments. Further, as we find that both IETS amplitude and shape is well converged with 4 vibrating unit cells, we fix the dynamical region to this size in the following analysis.

The computed IETS signals for the AGNR as a function of varying gate voltage are shown in \Figref{fig:IETSclean}(a) as a density plot. Specific IETS spectra at selected gate voltages are shown in \Figref{fig:IETSclean}(c)
for both the intrinsic part (temperature broadening at $T=4.2$~K) as well as that one would observe employing the experimental lock-in technique (additional broadening due to a modulation voltage of $V_\mathrm{rms}=5$~mV).
We find that for the AGNR there are generally two well-defined IETS signals
 appearing around $169$ and $196$ meV, corresponding to the D- (ring breathing) and G- (E$_{2g}$ phonon) modes, respectively, also observed in Raman spectroscopy.\cite{Han2007,Ferrari2013}. 
The D-signal also has a shoulder with a local maximum at $159$ meV with contributions from several modes. These three distinct features are indicated with vertical lines in \Figref{fig:IETSclean}(a,c). 
Shifting $E_F$ inside the gap region with a relatively small gate voltage $|V_G|\lesssim0.5$ V does not affect the IETS appreciably. However, when $E_F$ comes close to the conduction band of the AGNR the signal increases by a factor of five and a small peak-dip feature appear similar to the one reported for gated benzene-dithiol molecular contacts.\cite{Song2009,Lu2014} Upon further gating into the conduction band the IETS signals undergo a sign reversal (from peaks to dips) as the transmission increases beyond approximately $0.5$ for the involved channels.\cite{Paulsson2008} Similar effects are also found by gating into the valence band of the AGNR.

We can easily identify the most important vibrational mode vectors $v_\lambda$ 
for the IETS from the 
two amplitudes $\left|\gamma_\lambda\right|$ and $\left|\kappa_\lambda\right|$ given in Eqs.~(\ref{eq:gamma})-(\ref{eq:kappa}). 
These modes can further be analyzed in terms of the phonons in the infinite AGNR.
To do so we introduce the measure $F_{nk}$ representing the overlap between modes in the finite dynamical cell and the phonon band modes weighted by the size of the IETS signal,
\begin{equation}
    F_{nk}(V_G)=\sum_{\lambda}\left|\gamma_\lambda\right(V_G)|\left|u_{nk}\left(1,e^{ik},\ldots,e^{i(N-1)k}\right)\cdot v_{\lambda}\right|^2,  
    \label{eq.weightF}
\end{equation}
where $u_{nk}$ is the phonon band mode indexed by $n$, and $v_{\lambda}$ is the modes in a finite $N$ primitive cell long dynamical region index by $\lambda$. 

The projections $F_{nk}(V_G=0\mathrm{V})$ are depicted as widths of the phonon bands in \Figref{fig:AGNRphononband}(a),
where the red, green and blue colors refer to  modes with frequencies in the ranges $\hbar\omega>180$ meV, $180>\hbar\omega>162$ meV, and $\hbar\omega<162$ meV, respectively.
In total four bands contribute to the IETS signal corresponding to the four signals seen in the intrinsic part of the IETS spectrum in \Figref{fig:IETSclean}(c).
The corresponding $\Gamma$-point phonon modes inside the primitive cell for the infinite ribbon are shown in \Figref{fig:AGNRphononband}(b-e).

\subsection{\label{subsec:CleanZGNR} Pristine zigzag nanoribbon}

\begin{figure} 
\includegraphics[trim= 0cm 0cm 0cm 0cm,clip=true,width=1.\columnwidth]{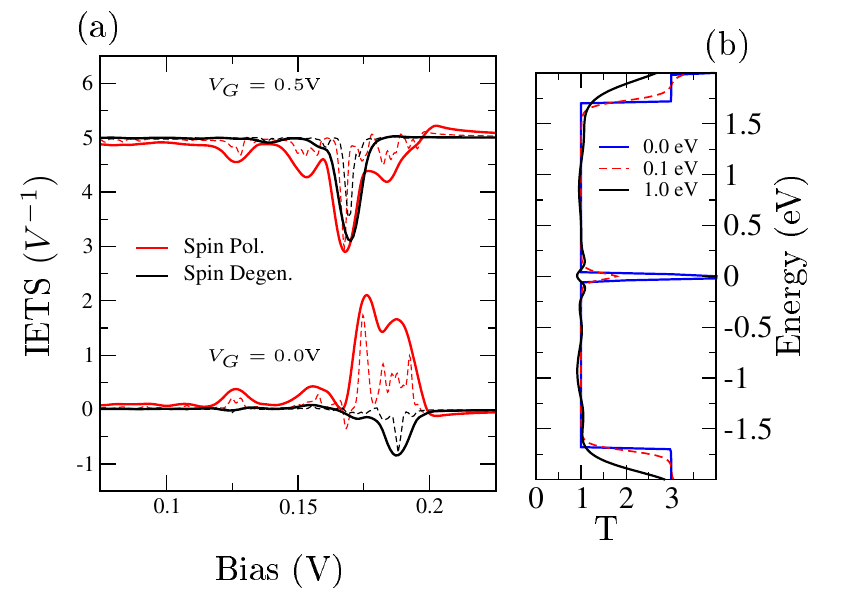}\vspace{-0.3cm}
\caption{\label{fig:NOspin} (Color online) (a) IETS signals for the pristine ZGNR (6 vibrating unit cells). The black lines correspond to spin-degenerate calculations while the red lines are the spin-up components of spin-polarized calculations. Broadening originates from temperature $T=4.2$~K
and modulation voltage $V_\mathrm{rms}=5$~mV (full lines) or $V_\mathrm{rms}=0$~mV (dashed lines). 
(b) Electronic transmission from spin-degenerate calculations with varying electrode broadening describing the coupling to the metal contacts, $\eta=0,0.1,1$~eV (see also \Figref{fig:CleanGNR}(g) for the corresponding spin-polarized case). }
\end{figure}

\begin{figure} 
    \includegraphics[trim= 0cm 0cm 0cm 0cm,clip=true,width=.98\columnwidth]{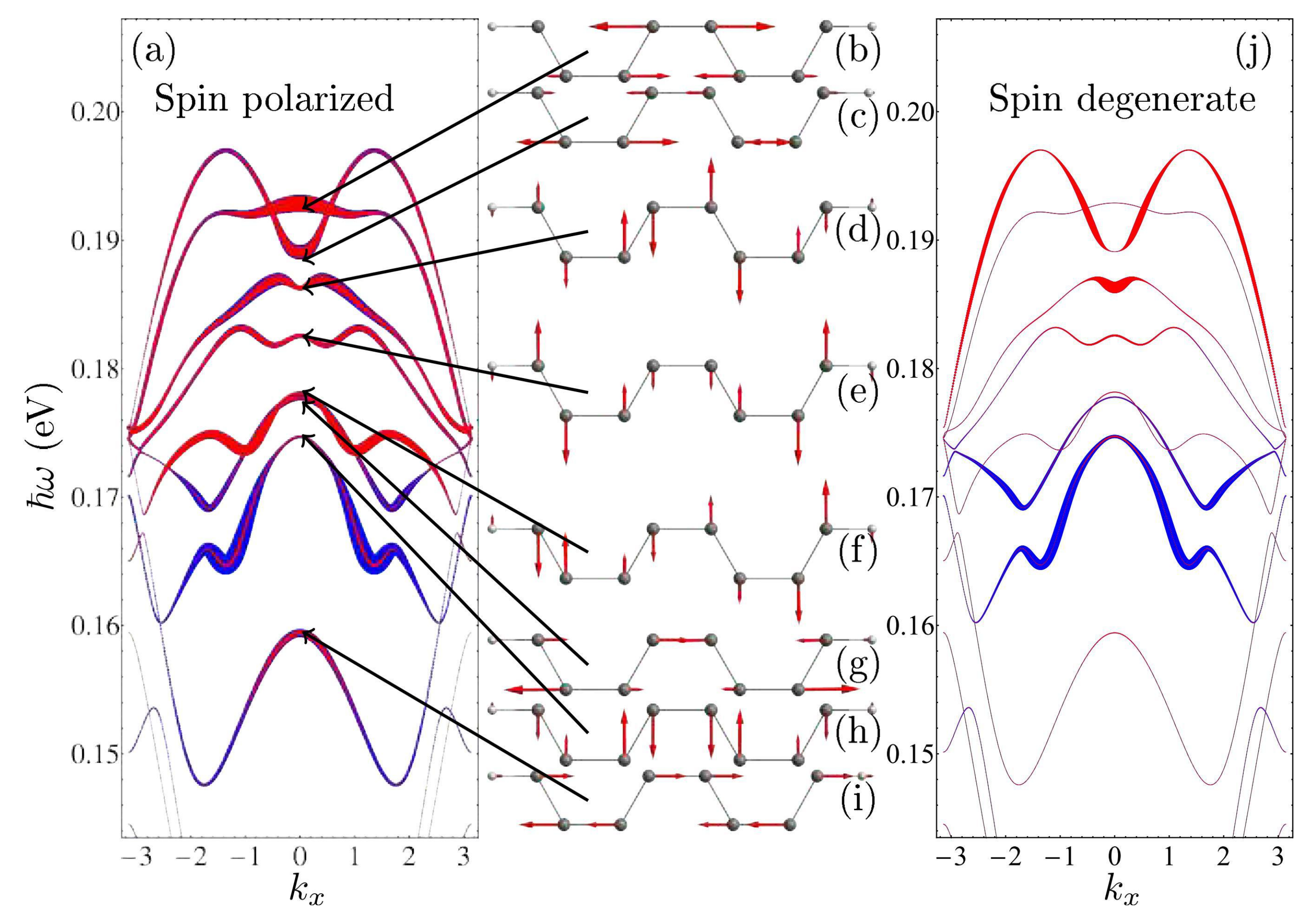}
    \caption{\label{fig:bandZGNR} (Color online) The phonon band structure of the ZGNR, ($k_x$ is in units of inverse unit cell length), together with the  $\Gamma$-point modes. The widths of the red bands are proportional to the weight function $F(0\mathrm{V})$ (\Eqref{eq.weightF}), while the widths of the blue bands are proportional to $F(0\mathrm{V})+F(0.5\mathrm{V})$.}
\end{figure}

\begin{figure*} 
\centering
\includegraphics[trim= 0cm 0cm 0cm 0cm,clip=true,width=1.8\columnwidth]{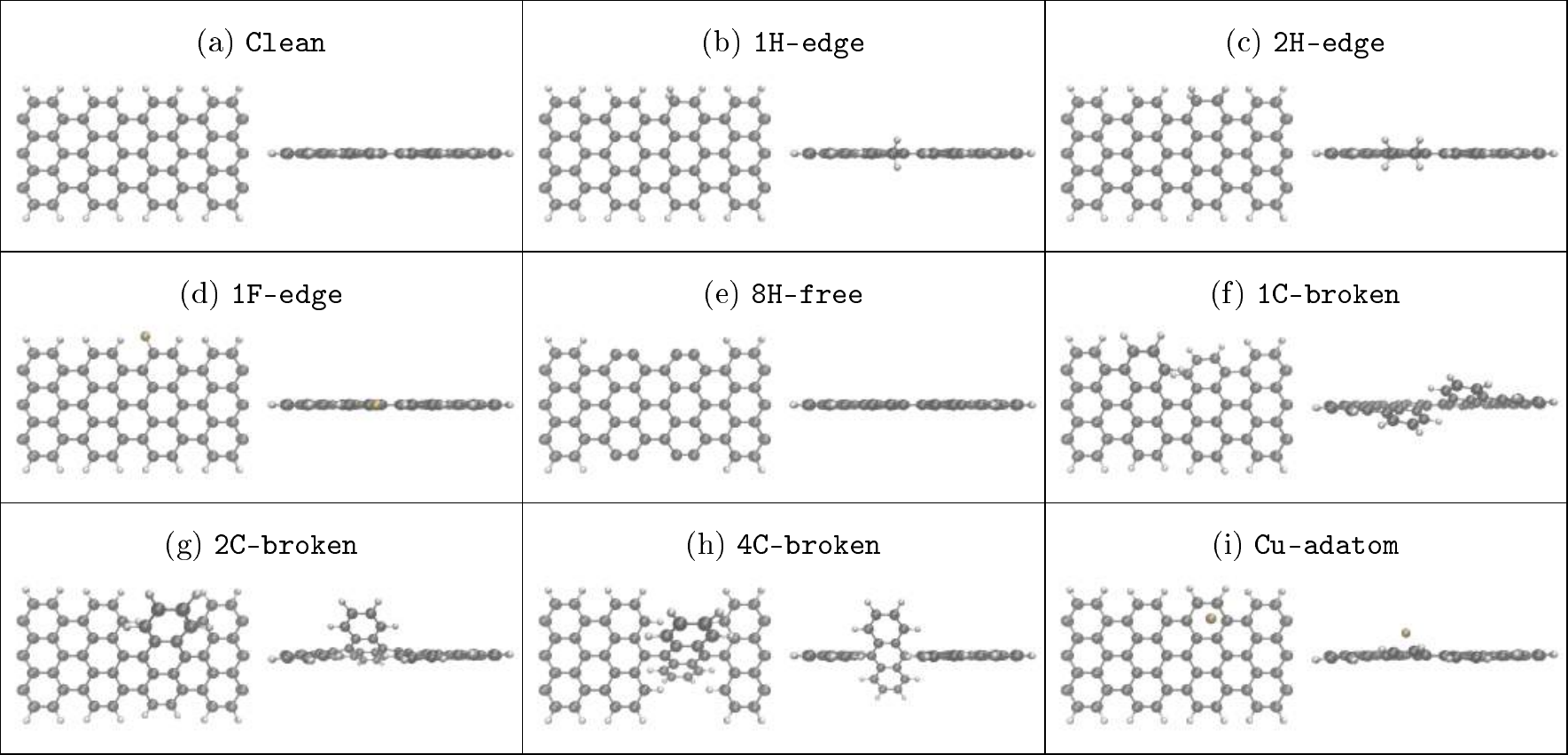}
\caption{(Color online)  Top and side views of the dynamical region describing the various AGNR defect structures. (a) Pristine AGNR. (b) One extra H atom on one of the edges. (c) Two extra H atoms on one of the edges. (d) One H atom replaced by a F atom. (e)  Dehydrogenated edge where 4 H atoms have been removed from each side. (f) One broken C-C bond. (g) Two broken C-C bonds. (h) Four broken C-C bonds. (i) Cu adatom in a hollow site on the edge.
}
\label{fig:xyzAGNR}
\end{figure*}

We next turn to our results for the pristine ZGNR shown in \Figref{fig:CleanGNR}(e). It has a width of $W=4$ zigzag ``chains'' (4-ZGNR) corresponding to a C-C edge distance of 7.26~\AA. The breaking of sublattice symmetry for the ZGNR and lack of pseudo-phase result in different selection rules for the matrix elements and difference in for example Raman signals.\cite{Saito2010a} The ZGNR generally presents spin-polarized edge states exhibiting a small band gap at the DFT level,\cite{Raza2011} in our case $E_g\approx 0.6$ eV (we note that this gap disappears in simpler tight-binding descriptions\cite{Raza2011} or spin-degenerate DFT calculations).
The spin-polarized edge states play the major role for the conduction, see the spin-down eigenchannels visualized in \Figref{fig:channels}(e-h).

Since the edge states break the mirror symmetry with respect to the middle of the ribbon, there are fewer symmetry-forbidden inelastic transitions between the scattering states for the ZGNR. Thus, we expect a wider range of modes to contribute to the IETS signal as compared to the AGNR case. Indeed this is in agreement with the findings shown in \Figref{fig:ConvClean}(b) and \Figref{fig:IETSclean}(b,d). 
The greater number of modes contributing to the IETS for the ZGNR results in broader signals with similar magnitudes as compared to the IETS for AGNR. 
As for the AGNR case the IETS signal is well converged with a dynamical region consisting of 6 vibrating unit cells [\Figref{fig:ConvClean}(b)].

For ZGNRs the ring breathing is forbidden by symmetry, thus the IETS is generally characterized by transverse and longitudinal modes.
To explore the impact of spin-polarization on the ZGNR-IETS we compare in \Figref{fig:NOspin} the results from both spin-degenerate and spin-polarized calculations. Without gate voltage ($V_G=0$ V) the IETS display opposite signs due to the spin-induced gap. Only a single peak contributes to the spin-degenerate IETS while several peaks contribute to the spin-polarized IETS. Even if the ZGNR is tuned by $V_G=0.5$V to become metallic and the two treatments then show the same overall sign in IETS, the spin-polarized IETS persists to show a much richer structure. This difference suggests that IETS could be a way to indirectly observe spin-polarized edge states.

Projecting the modes contributing to the IETS onto the phonon band modes further underlines how several bands with different symmetries contribute to the spin-polarized IETS, while only a couple of bands contributes to the spin-degenerate IETS, see \Figref{fig:bandZGNR}. Again we use \Eqref{eq.weightF} for this characterization, where the overlap for $V_G=0.0$ V corresponds to the red color and the overlap for $V_G=0.5$ V corresponds to the difference between the blue and red color in \Figref{fig:bandZGNR}, respectively.
It is clear that spin-polarization permits more modes to contribute to the IETS.
In contrast to the spin-degenerate case, where the symmetric electronic states (with respect to the middle of the ribbon) only can couple to the symmetric vibration modes, the symmetry lowering of the electronic states by spin-polarization opens up also for scattering also via odd modes.

\section{Defected graphene nanoribbons}

In this section we address the modification and new signals in IETS that arise due to various defects in the GNR. Regardless of the fabrication method, defects will inevitable occur. For example, if the AGNRs are synthesized from a precursor molecule, involving heating and dehydrogenation, as reported by Cai \etal\cite{Cai2010} and Blankenburg \etal,\cite{Blankenburg2012} there is a chance that the reaction is incomplete and some of the C-C bonds between the precursor molecules do not form. Also there is a chance that a part of the final AGNR will have dehydrogenated edges or are passivated by two hydrogen atoms. Finally, defects may be introduced on purpose by locally dosing a high current from the tip of a STM.\cite{VanderLit2013f}

\subsection{\label{subsec:DefectAGNR} Defects in AGNRs}

\begin{figure} 
\includegraphics[clip=true,width=\columnwidth]{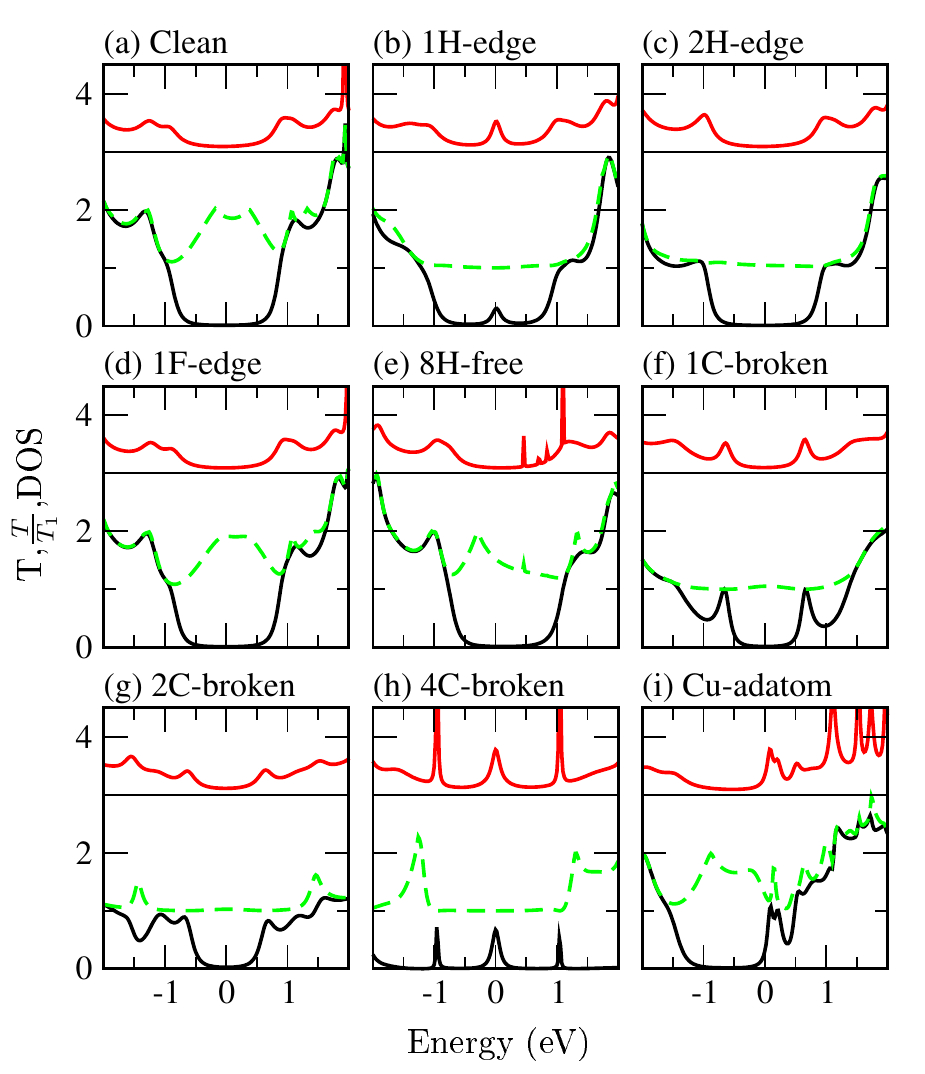}
\caption{\label{fig:TransAGNR} (Color online) Electronic properties of the AGNR structures shown in \Figref{fig:xyzAGNR}. The total transmission is shown with black lines. The ratio ${T}/{T_1}$, where $T_1$ is the transmission originating from the most transmitting eigenchannel is shown with green dashed lines (this ratio gives a lower bound to the number of contributing eigenchannels). The DOS for the C atoms in the dynamical region is shown with red lines (offset by 3 units).}
\end{figure} 

\begin{figure*} 
\includegraphics[clip=true,width=1.8\columnwidth]{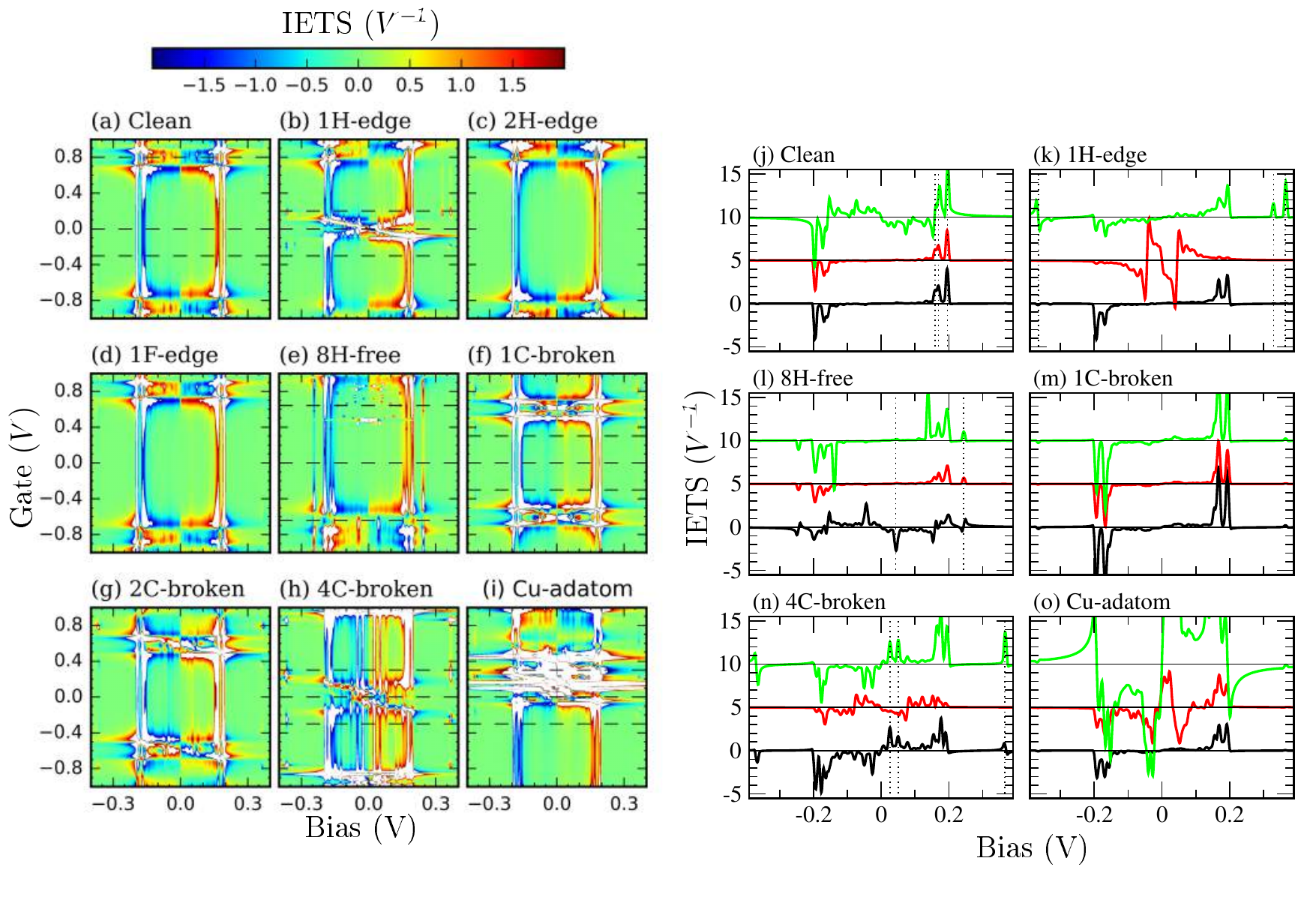}\vspace{-1cm}
\caption{\label{fig:IETSdefectAGNR} (Color online) (a-i) IETS as a function of gate voltage $V_G$ for the pristine and defected AGNR structures shown in \Figref{fig:xyzAGNR} . 
(j-o) IETS for six selected structures at three specific gate values (dashed horizontal lines in panels a-i). The curves are offset with the most negative gate value at the bottom (black curves) and the most positive at the top (green curves). 
(j) Clean AGNR at gate values $V_G=-0.3$, 0.0, and 0.8~V. (k) \texttt{1H-edge} at $V_G=-0.3$, 0.0, and 0.2~V. (l) \texttt{8H-free} at $V_G=-0.3$, 0.0, and 0.6~V. (m) \texttt{1C-broken} at $V_G=-0.3$, 0.0, and 0.3~V. (n) \texttt{4C-broken} at $V_G= -0.3$, 0,0, and 0.3~V. (o) \texttt{Cu-adatom} at $V_G= -0.3$, 0.0, and 0.3~V. Dotted vertical lines are guides to the eye of characteristic IETS signals corresponding to the modes in \Figref{fig:ArmModes}}.
\end{figure*}

\begin{figure*} 
 \includegraphics[clip=true,width=2\columnwidth]{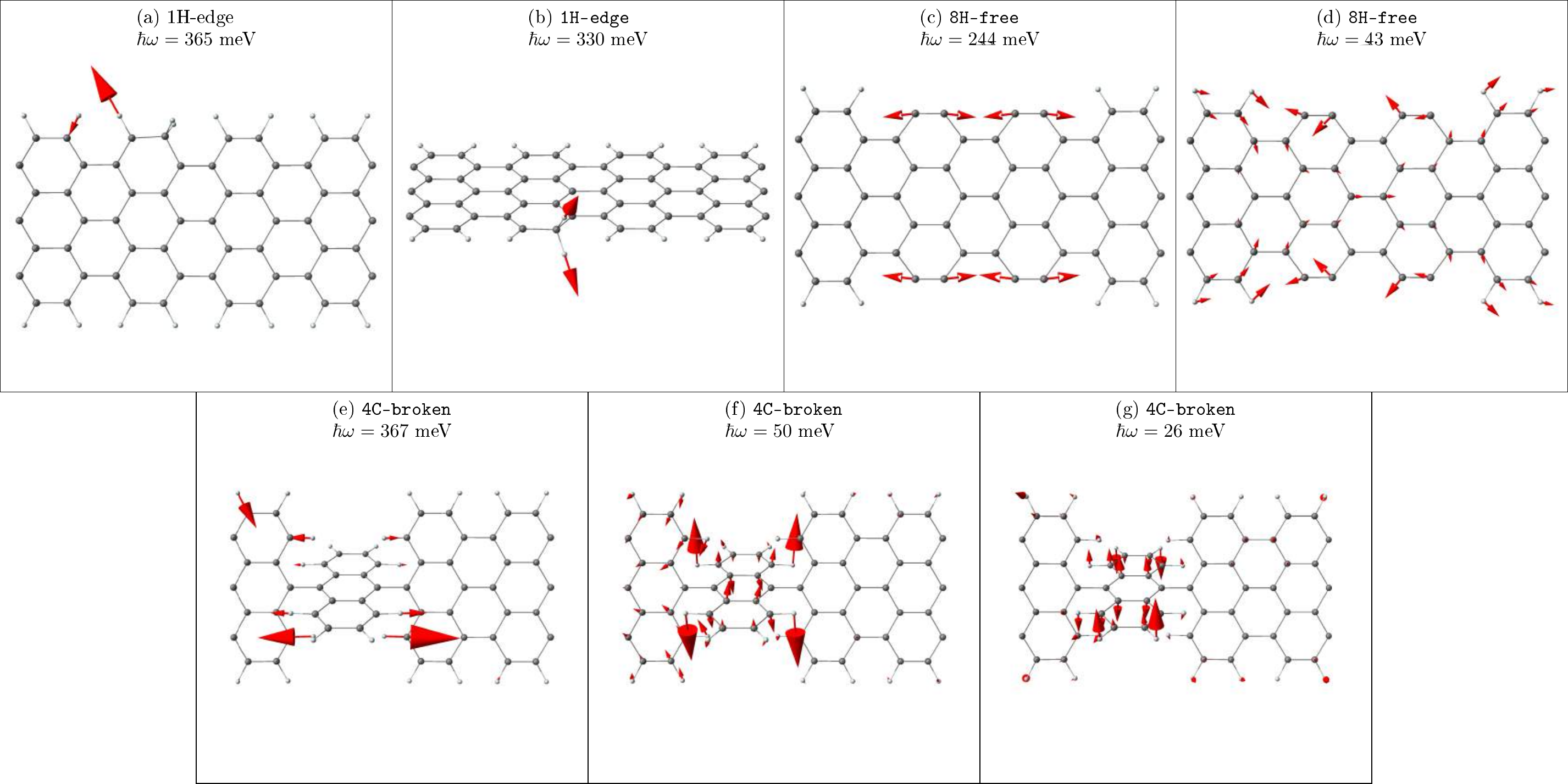}
 \caption{\label{fig:ArmModes} (Color online) Visualization of the  most contributing defect-induced vibrational modes to the IETS signals indicated by vertical lines in \Figref{fig:IETSdefectAGNR}(j-o).
(a-b) The two hydrogen signals for \texttt{1H-edge}. (c) Localized edge mode at the carbon dimers for the \texttt{8H-free}. (d) Delocalized edge mode for the \texttt{8H-free}. (e) Hydrogen mode from the zigzag edge of \texttt{4C-broken}. (f-g) Defect modes for \texttt{4C-broken}.
}
\end{figure*}

In \Figref{fig:xyzAGNR} we show the structures of pristine AGNR along with 8 different defect configurations which we have considered. These include four defects in the edge passivation as follows: A single edge side with an extra hydrogen atom [\texttt{1H-edge}, \Figref{fig:xyzAGNR}(b)], two edge sides with each an extra hydrogen atom [\texttt{2H-edge}, \Figref{fig:xyzAGNR}(c)], one hydrogen replaced by a fluorine atom [\texttt{1F-edge}, \Figref{fig:xyzAGNR}(d)], and a dehydrogenated edge with 4 hydrogen atoms removed from each side  [\texttt{8H-free}, \Figref{fig:xyzAGNR}(e)]. 
We have also considered defects in the atomic structure in the form of one, two, or four broken C-C bonds [\texttt{1C-broken}, \texttt{2C-broken}, \texttt{4C-broken}, \Figref{fig:xyzAGNR}(f)-(h)] as well
as a Cu adatom on the AGNR [\texttt{Cu-adatom}, \Figref{fig:xyzAGNR}(i)]. 
For all these systems the entire dynamical region was relaxed, i.e., the parts of the AGNRs shown in \Figref{fig:xyzAGNR}. 

Defects may influence the IETS signal in two ways. First, a defect can have a direct impact by changing the vibrational degrees of freedom. In order for the change in the vibrational spectrum to give a signal in the IETS, the new vibrations must couple to the current, and preferably have frequencies which do not coincide with ones already giving IETS signals for the pristine ribbons.
Second, a defect can substantially change the electronic structure and thereby have an impact on the e-ph couplings associated with the active modes or even the transmission eigenchannels of the pristine ribbons, e.g., changing a peak in the IETS to a dip (and vice versa) or enhancing asymmetric contributions via \Eqref{eq:asym}.          

The electronic properties of the pristine AGNR is shown in \Figref{fig:TransAGNR}(a). The carbon DOS projected to the device region (red curve) reveals a gap as expected from the band structure [\Figref{fig:CleanGNR}(b)], which is significantly broadened from the coupling to the metallic electrodes. The two valence and two conduction bands in the considered energy range naturally explain that the total transmission (black curve) is bound below a value of 2.
Further, the ratio ${T}/{T_1}<2$ (green dashed line), measuring the minimum number of contributing channels where $T_1$ is the transmission of the most transmitting eigenchannel, shows that both channels play a role for the transport, at least away from the edges of the direct band gap. Measurements of shot noise may provide insights into this effective number of conductance eigenchannels.\cite{RBG_Djukic_single_molecule,c60} 
We can now discuss how the different defects modify the electronic properties.
From \Figref{fig:TransAGNR}(b)-(i) we notice that not all defects change the elastic transmission, and furthermore, a change in elastic transmission needs not be unique for a specific defect.

Instead, IETS may provide a additional fingerprint in the current that can be used to identify the type of defect. Figure \ref{fig:IETSdefectAGNR} shows the computed IETS as a function of gate voltage for the 8 different defects. As for the clean structure, the two peaks at $169$ and $196$ meV corresponding to the D- and G- Raman modes are dominant for a range of gate values for all the structures. Another feature, which is present in all the systems, is the appearance of several signals close to the band onsets.
In the following subsections we discuss in more detail the transport characteristics with the different types of defects in AGNRs.

\subsubsection{Edge passivation}

Considering defects in the edge passivation [\Figref{fig:xyzAGNR}(b-e)] the gap in the transmission is essentially unchanged [\Figref{fig:TransAGNR}(b-e)], except for the \texttt{1H-edge} structure where a zero-energy resonance appears in the DOS and transmission [\Figref{fig:TransAGNR}(b)]. This new peak can be attributed to tunneling via a mid-gap state which appears due to the local breaking of sub-lattice symmetry.\cite{Raza2011} Thus, if a H atom is added to the neighboring C atom [\texttt{2H-edge}, \Figref{fig:xyzAGNR}(c)] the peak disappears [\Figref{fig:TransAGNR}(c)]. The addition of one or two H atoms on the same side also results in the closing of one transmission channel between the valence and conduction bands as shown in \Figref{fig:TransAGNR}(b,c). Concerning the vibrational degrees of freedom,  the addition of extra hydrogen to the edge results in new vibrational modes around 330~meV for \texttt{1H-edge} and around 343 and 353~meV for \texttt{2H-edge}, clearly outside the bulk phonon band (ranging up to $\sim 200$ meV) of pristine AGNR.\cite{madse10}

Comparing the IETS in \Figref{fig:IETSdefectAGNR}(a-c) we find that only \texttt{1H-edge} gives a signal which differs significantly from the pristine case. Figure \ref{fig:IETSdefectAGNR}(k) shows
specific IETS for selected gate voltages for \texttt{1H-edge}. Here, at $V_G=0.2$V (top green curve) we see how new signals appear at large voltages: For positive bias polarity two signals appear at 330 and 365~meV, respectively, while for negative bias polarity only an asymmetric signal around $-365$~meV is present. The signal at 330~meV is due to vibrations of the $\rm{H}_2$ [\Figref{fig:ArmModes}(b)], while the signal at 365~meV [\Figref{fig:ArmModes}(a)] is due to the H atom on the neighboring C atom. Further, the amplitude of the signals around 169 and 196 meV is also found to depend on bias polarity.

Gating onto the zero-energy resonance for \texttt{1H-edge} the IETS signal [middle red curve in \Figref{fig:IETSdefectAGNR}(k)] is dominated by large asymmetric signals for low energy vibrations due to the contribution from $\kappa_\lambda$ and \Eqref{eq:asym}. We note that $\kappa_\lambda$ changes sign with bias polarity for this approximately left-right symmetric structure. This can be seen from the red IETS curve in \Figref{fig:IETSdefectAGNR}(k) which is roughly an odd function of the bias voltage. In close proximity of the zero-energy resonance a characteristic ``X-shape'' is observed in the gate-dependent IETS, while away from it the signals approach that of the pristine AGNR [\Figref{fig:IETSdefectAGNR}(b)].  
 
Substituting a H atom with a F atom (\texttt{1F-edge}) is seen to have virtually no effect in the IETS of \Figref{fig:IETSdefectAGNR}(d). This suggests that a significant change in the chemical composition directly involving the $\pi$-electronic system is required in order to obtain a signal although the vibrations are influenced by the heavier passivation. 

Such a significant change in the passivation occurs for instance by removing four H atoms on each side (\texttt{8H-free}), giving rise to four very narrow peaks in the DOS around the conduction band, [\Figref{fig:TransAGNR}(e)]. These correspond to very localized dangling-bond states on the dehydrogenated dimers and therefore do not show up in the transmission.
However, the dehydrogenated edges give rise to localized vibrations outside the range of the pristine vibrational spectrum.\cite{madse10} The in-phase vibration of the dehydrogenated C dimers at the armchair edges [\Figref{fig:ArmModes}(f)] gives rise to an extra IETS peak at 244~meV [\Figref{fig:IETSdefectAGNR}(l)]  matching the H-free mode measured by Raman.\cite{Huang2012b} We find that this signal is robust as it appears in the whole range of gate values. When gating into to the valence band a new signal appears around 43~meV [$V_G\approx -0.8$ V in \Figref{fig:IETSdefectAGNR}(e)] originating from a low energy edge vibration [\Figref{fig:ArmModes}(g)].

\subsubsection{Structural defects}

The electronic transmission in GNRs is mediated by the carbon $\pi$ system. Thus if a C-C bond fails to be formed during GNR synthesis or if it is broken again at a later stage, a large effect can be expected for the electronic conduction properties. This impact is indeed revealed in \Figref{fig:TransAGNR}(f-h).
Breaking one or two bonds results in the formation of two in-gap states which, broadened by the electrodes, make the gap appear smaller. 
The IETS signals for the \texttt{1C-broken} and \texttt{2C-broken} in \Figref{fig:IETSdefectAGNR}(f,g,m) have the same two signals at $169$ and $196$ meV as for the clean ribbon. However, the relative amplitudes are interchanged such that the "D"-peak is now slightly more intense than the "G"-peak.

Breaking four C-C bonds [\texttt{4C-broken}, \Figref{fig:xyzAGNR}(h)], resulting in constrictions of single C-C bonds, totally alter the DOS which is now dominated by three sharp peaks as seen in \Figref{fig:TransAGNR}(h).
The corresponding IETS signals are shown in \Figref{fig:IETSdefectAGNR}(h,n). In the proximity of the zero-energy resonance a broad range of signals at low vibrational energies appears (red curve in panel n) as well as a characteristic ``X-shape'' in the gate plot (panel h) similar to that of \texttt{1H-edge}.
Gating away from the resonance we observe two additional robust IETS signals at $27$ and $50$~meV resulting from vibrations localized at the defect [\Figref{fig:ArmModes}(d,e)].

\subsubsection{Adatoms}
Transition metals are typically used for growth of graphene or as a substrate for the bottom-up synthesis of GNRs. Thus it is of interest to consider the effect of adatoms of this type on GNRs. A Cu adatom on graphene adsorbs preferentially in the on-top position.\cite{Liu2012} However, positioning Cu such that it breaks the axial symmetry of our AGNR, we find that it is most stable in a hollow site at the edge [\texttt{Cu-adatom}, \Figref{fig:xyzAGNR}(i)]. The DOS and transmission in \Figref{fig:TransAGNR}(i) reveal a n-type doping effect shifting $E_F$ close to the conduction band while leaving the two transmission channels inside the gap relatively intact.

For the pristine GNR the e-ph couplings of the out-of-plane vibrations are suppressed due to the symmetry of the $\pi$-orbitals. However, around the onset of the conduction band the IETS signals in  \Figref{fig:IETSdefectAGNR}(i,o) is dominated by large asymmetric signals with significant contributions from out-of-plane phonons. These modes come into play due to breaking of the planar symmetry by the adatom. Also note that by gating of $E_F$ within the gap these signatures of the adatom disappear, cf.~the lower black curve in \Figref{fig:IETSdefectAGNR}(o).

\subsection{\label{subsec:ZGNR} Defects in ZGNRs}

\begin{figure*} 
 \includegraphics[trim= 0cm 0cm 0cm 0cm,clip=true,width=1.8\columnwidth]{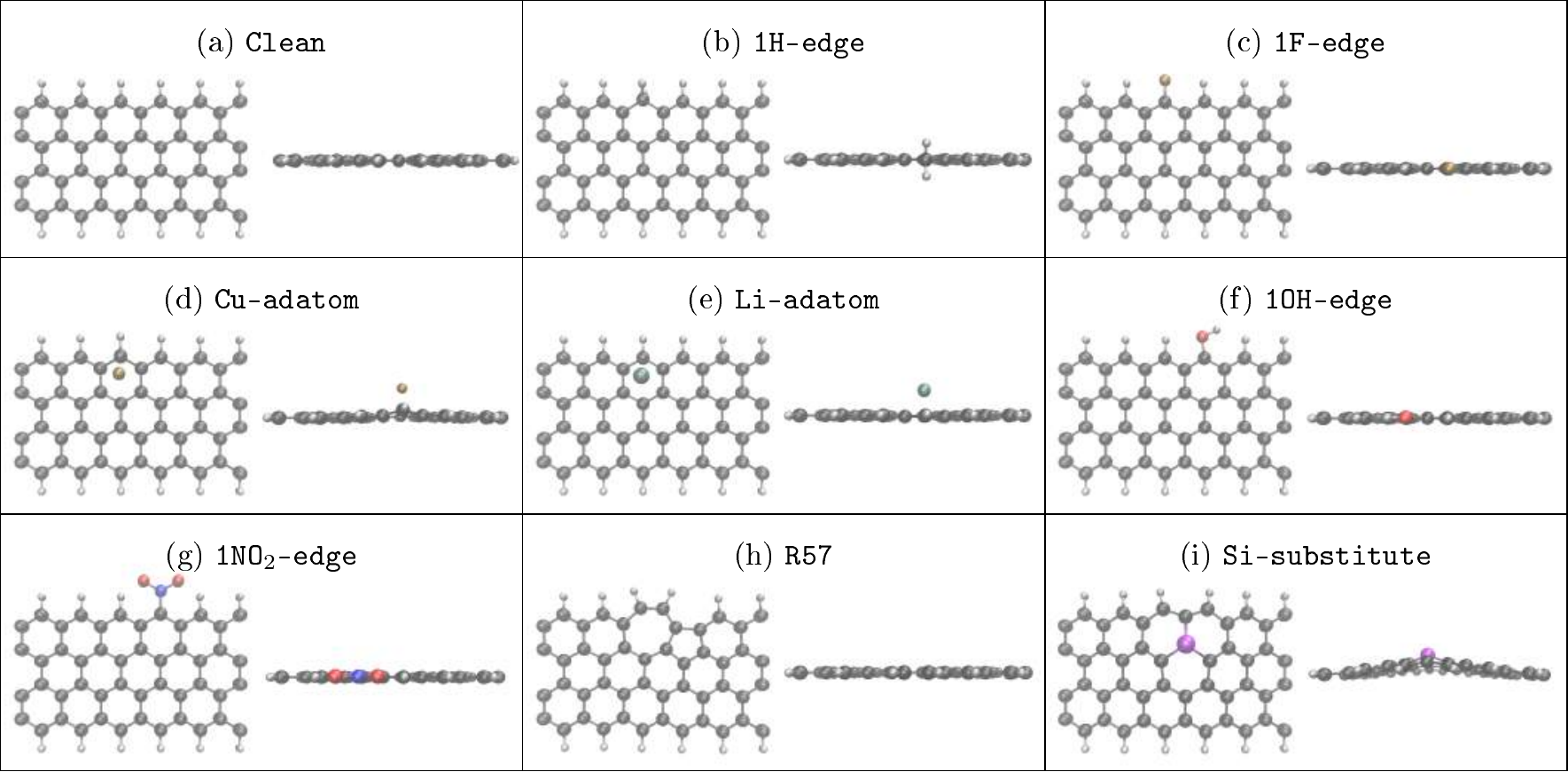}
   \caption{\label{fig:StructDefectZGNR}
  (Color online) Top and side views of the dynamical region describing the various ZGNR defect structures. (a) Pristine ZGNR. (b) One extra H atom on one of the edges. (c) One H atom replaced by a F atom. (d) Cu adatom in a hollow site on the edge. (e) Li adatom in a hollow site on the edge. (f) One H replaced by a OH group. (g) One H replaced by a $\rm{NO}_2$ group. (h) Structural defect (\texttt{R57}). (i) Substitutional Si defect next to the edge.
  }  
\end{figure*}
 
\begin{figure*} 
   \includegraphics[trim= 0cm 0cm 0cm 0.cm,clip=true,width=1.85\columnwidth]{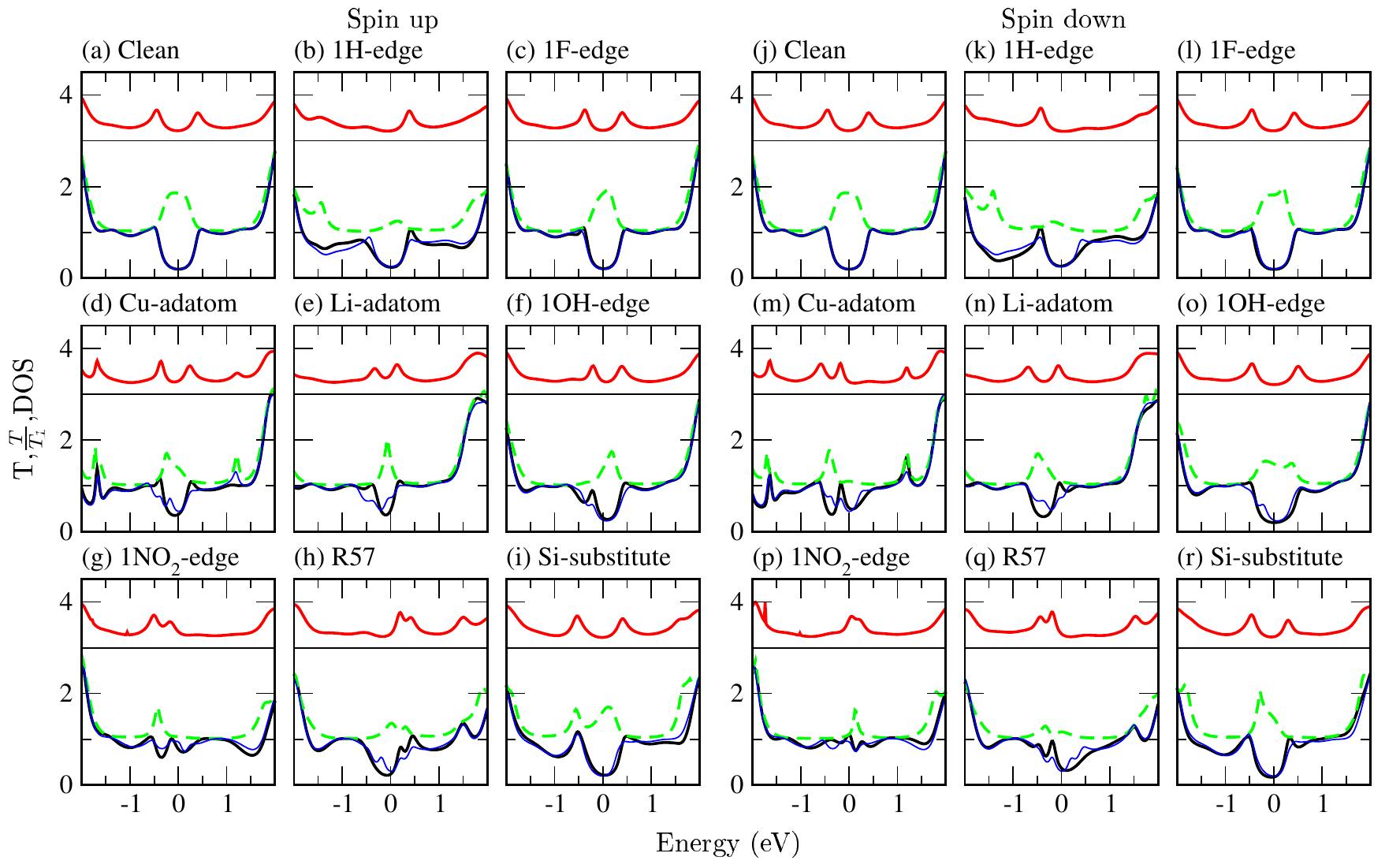}\vspace{-0.3cm}
    \caption{\label{fig:TransDefectZGNR}
   (Color online) Electronic properties of the ZGNR structures shown in \Figref{fig:StructDefectZGNR} with the spin-up/down components in the left/right panel. 
   The spin-resolved total transmission is shown with black lines 
   while spin-averaged total transmission is shown with thin blue lines.
   The ratio ${T^\sigma}/{T_1^\sigma}$, where $T_1^\sigma$ is the transmission originating from the most transmitting spin eigenchannel, is shown with green dashed lines (this ratio gives a lower bound to the number of contributing eigenchannels with spin $\sigma$). The spin-resolved DOS for the C atoms in the dynamical region is shown with red lines (offset by 3 units).
   }
\end{figure*}

Let us next consider a series of defects for the zigzag graphene nanoribbon.
In \Figref{fig:StructDefectZGNR} we show the atomic structures of pristine ZGNR along with 8 different defect configurations. We consider the following defects in the edge-passivation: A single edge with an extra hydrogen [\texttt{1H-edge}, \Figref{fig:StructDefectZGNR}(b)], one hydrogen is replaced by either a F atom [\texttt{1F-edge}, \Figref{fig:StructDefectZGNR}(c)], an OH group [\texttt{1OH-edge}, \Figref{fig:StructDefectZGNR}(f)], or a $\rm{NO}_2$ group [\texttt{1NO$_2$-edge}, \Figref{fig:StructDefectZGNR}(g)].   
We also consider defects in the form of a Cu adatom [\texttt{Cu-adatom}, \Figref{fig:StructDefectZGNR}(d)] or a Li adatom [\texttt{Li-adatom}, \Figref{fig:StructDefectZGNR}(e)]. Finally, we also study the effect of a structural defect in form of a 57 reconstruction [\texttt{R57}, \Figref{fig:StructDefectZGNR}(h)] and a substitutional defect where a C atom next to the edge is replaced by a Si atom [\texttt{Si-substitute}, \Figref{fig:StructDefectZGNR}(i)].
For all these systems the entire dynamical region was relaxed, i.e., the parts of the ZGNRs shown in \Figref{fig:StructDefectZGNR} using spin-polarized treatments.
The spin degrees of freedom $\sigma=\uparrow,\downarrow$ generalizes $\gamma_\lambda^\sigma$ and $\kappa_\lambda^\sigma$ [Eqs.\eqref{eq:gamma}-\eqref{eq:kappa}] corresponding to two independent spin channels, which in general can have quite different amplitudes and even opposite sign. The observable IETS would simply be the sum of these two components $(\partial_V^2 I_\uparrow + \partial_V^2 I_\downarrow)/(\partial_V I_\uparrow + \partial_V I_\downarrow)$.

Similar to the AGNR case, the electronic properties in the device region with the different impurity configurations for the ZGNR, now spin resolved, are summarized in \Figref{fig:TransDefectZGNR}.
The IETS of pristine ZGNR was already discussed in \Secref{subsec:CleanZGNR} and below we continue describing the IETS fingerprints for the various defects.

\subsubsection{Edge passivation}

As commented above, the broader IETS signals of pristine ZGNR [Figs.~\ref{fig:ConvClean}-\ref{fig:IETSclean}] (as compared with AGNR) can be understood from the breaking of the axial mirror symmetry and hence fewer symmetry-forbidden inelastic transitions. These broader signals may in general make the detection of defect signatures more difficult. For \texttt{1H-edge} [\Figref{fig:TransDefectZGNR}(b)] the IETS resembles that of the pristine ZGNR [\Figref{fig:TransDefectZGNR}(a)] inside the gap.
However, gating into the valance band [black curve in \Figref{fig:TransDefectZGNR}(k)] the edge states start to extend into the middle of the ribbon, partially restoring mirror symmetry, and thus resulting in part of the pristine ZGNR signals to disappear.

Here an extra signal appear due to edge-modes in the frequency range 194 to 199~meV with the most contributing mode at 196~meV as shown in [\Figref{fig:ZigModes}(a)]. The resulting IETS signal can clearly be seen in the bottom curve in \Figref{fig:IETSdefectZGNR}(k). As for the AGNR substituting a hydrogen with a fluorine atom has a very limited effect on the electronic properties and the IETS signal.

Substituting a hydrogen with an OH group, according to \Figref{fig:TransDefectZGNR}(f) and (o), have only a small effect on the spin down electrons, while it shrinks and add additional structure to the gap for the spin up electrons. For the spin up electrons there is a small peak inside the gap which gives rise to a large asymmetric IETS signal around $V_G=-0.2$V in \Figref{fig:TransDefectZGNR}(o) lower curve, compared to the pristine case. The most contributing mode to the asymmetric IETS signal is shown in \Figref{fig:ZigModes}(b). However, there is no clear signature of the OH group itself. In the same manner the substitution with a $\rm{NO}_2$ group removes the gap in the electronic properties without leaving any direct fingerprint of the $\rm{NO}_2$ group in the IETS signal.

\subsubsection{Adatoms}
As for the AGNR we consider the effect of adatoms. 
For the Cu adatom the transport gap shrinks for the spin up electrons while there is an in-gap peak for the spin down electrons, \cf \Figref{fig:TransDefectZGNR}(m). Thus, for some gate values the IETS signals reflect that the spin down electrons will back scatter while the spin up electrons will be forward scattered, and the observed signal is then the sum of these contributions. For a gate value of $V_G=-0.2$ V, the IETS signal is dominated by spin down electrons. Due to the finite width of the in-gap peak, in the spin down transmission, the low frequency phonons ($\hbar\omega<0.1$~meV) give rise to back scattering while the high frequency phonons ($\hbar\omega>0.1$~meV)  result in forward scattering. Thus, the low and high energy signals have different signs as can be seen from \Figref{fig:IETSdefectZGNR}(l). Interestingly, the low energy signal primarily consists of symmetric contributions from out-of-plane modes [\Figref{fig:ZigModes}(c)]. Replacing the Cu adatom with Li, the transmission and DOS, shown in \Figref{fig:TransDefectZGNR}(e,n), reveals  a spin dependent n-type doping effect, where $E_F$ is shifted the most for spin down. However, no in-gap peak is seen as for Cu and the IETS show no clear signature of the Li atom.

\subsubsection{Structural defect}
 
The formation of a \texttt{R57} reconstruction results in peaks in the DOS in the device region, just above $E_F$ for spin up [\Figref{fig:TransDefectZGNR}(h)] and just below $E_F$ for spin down [\Figref{fig:TransDefectZGNR}(q)]. The \texttt{R57} breaks the symmetry both in the vibrational and electronic structure allowing for IETS signals from a wider range of vibrations, resulting in broader peaks, as seen from \Figref{fig:IETSdefectZGNR}(h) and \Figref{fig:IETSdefectZGNR}(n). One of the contributing modes is localized at the border between the pentagon-ring and middle of the ribbon at $\hbar\omega=204\rm{meV}$ [\Figref{fig:ZigModes}(d)]. This localized mode yield a relatively small signal compared to the other signals, however, contrary to the other modes the localized mode is not expected to be broadened if the coupling to phonons away from the dynamical region is taken into account. The breaking of symmetry in the electronic structure also give rise to difference signals for the two bias polarities.

\subsubsection{Substitutional impurity}
\begin{figure*} 
 \includegraphics[clip=true,width=1.8\columnwidth]{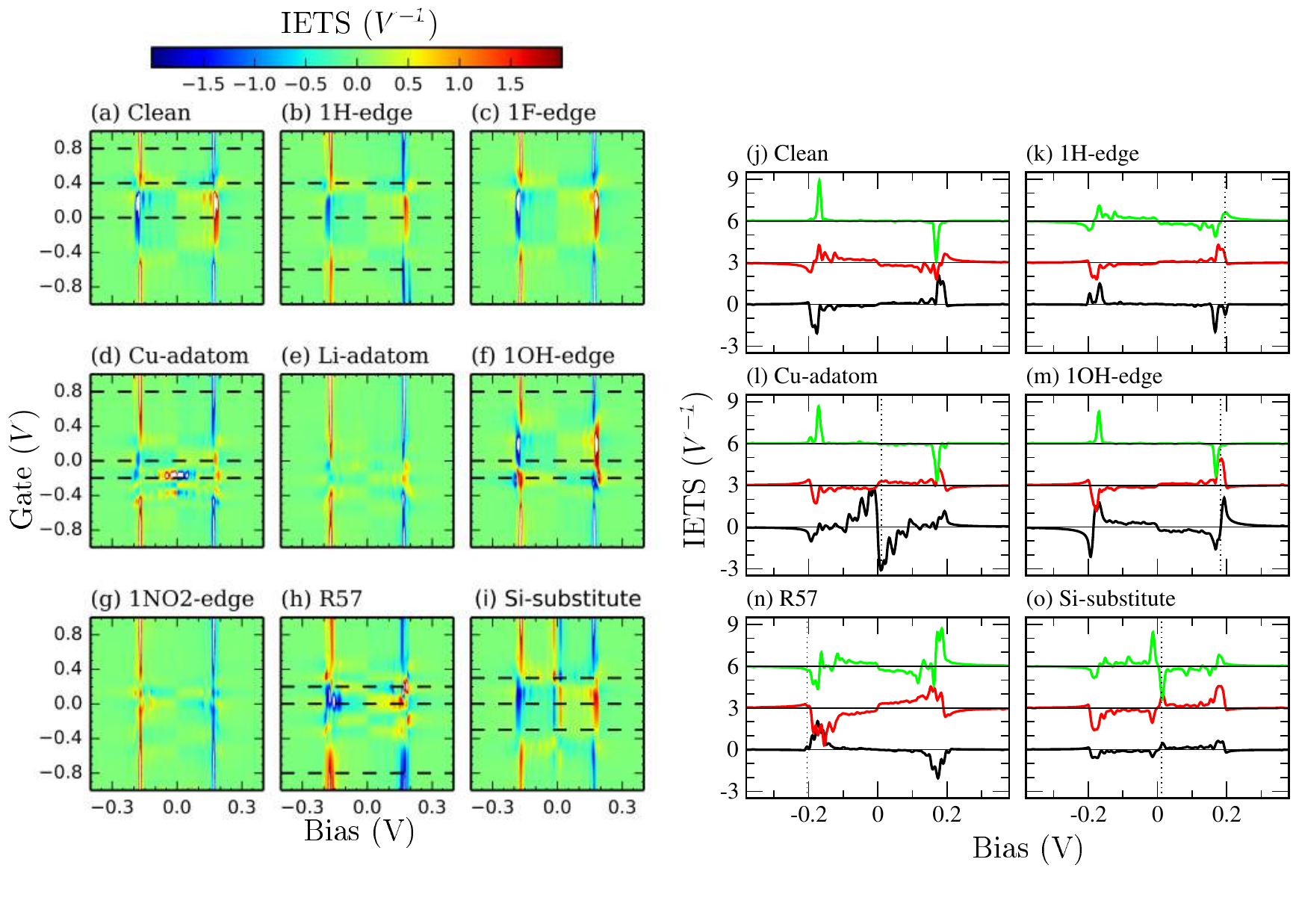}\vspace{-1cm}
 \caption{\label{fig:IETSdefectZGNR} (Color online) (a-i) Spin-averaged IETS as a function of gate voltage $V_G$ for the pristine and defected ZGNR structures shown in \Figref{fig:StructDefectZGNR}. 
  (j-o) IETS for six selected structures at three specific gate values (dashed horizontal lines in panels a-i). The curves are offset with the most negative gate value at the bottom (black curves) and the most positive at the top (green curves).
 (j) Clean ZGNR for gate values $V_G=0.0$, 0.4, and 0.8~V. (k) \texttt{1H-edge} at $V_G=-0.6$, 0.0, and 0.4~V. (l) \texttt{Cu-adatom} at $V_G= -0.2$, 0.0, and 0.8~V. (m) \texttt{1OH-edge} at $V_G=-0.2$, 0.0, and 0.8~V. (n) \texttt{R57} at $V_G=-0.8$, 0.0, and 0.2~V. (o) \texttt{Si-substitute} at $V_G=-0.3$, 0.0, and 0.3~V. Dotted vertical lines are guides to the eye of characteristic IETS signals corresponding to the modes shown in \Figref{fig:ZigModes}.}
 \end{figure*}

\begin{figure*} 
 \includegraphics[clip=true,width=1.6\columnwidth]{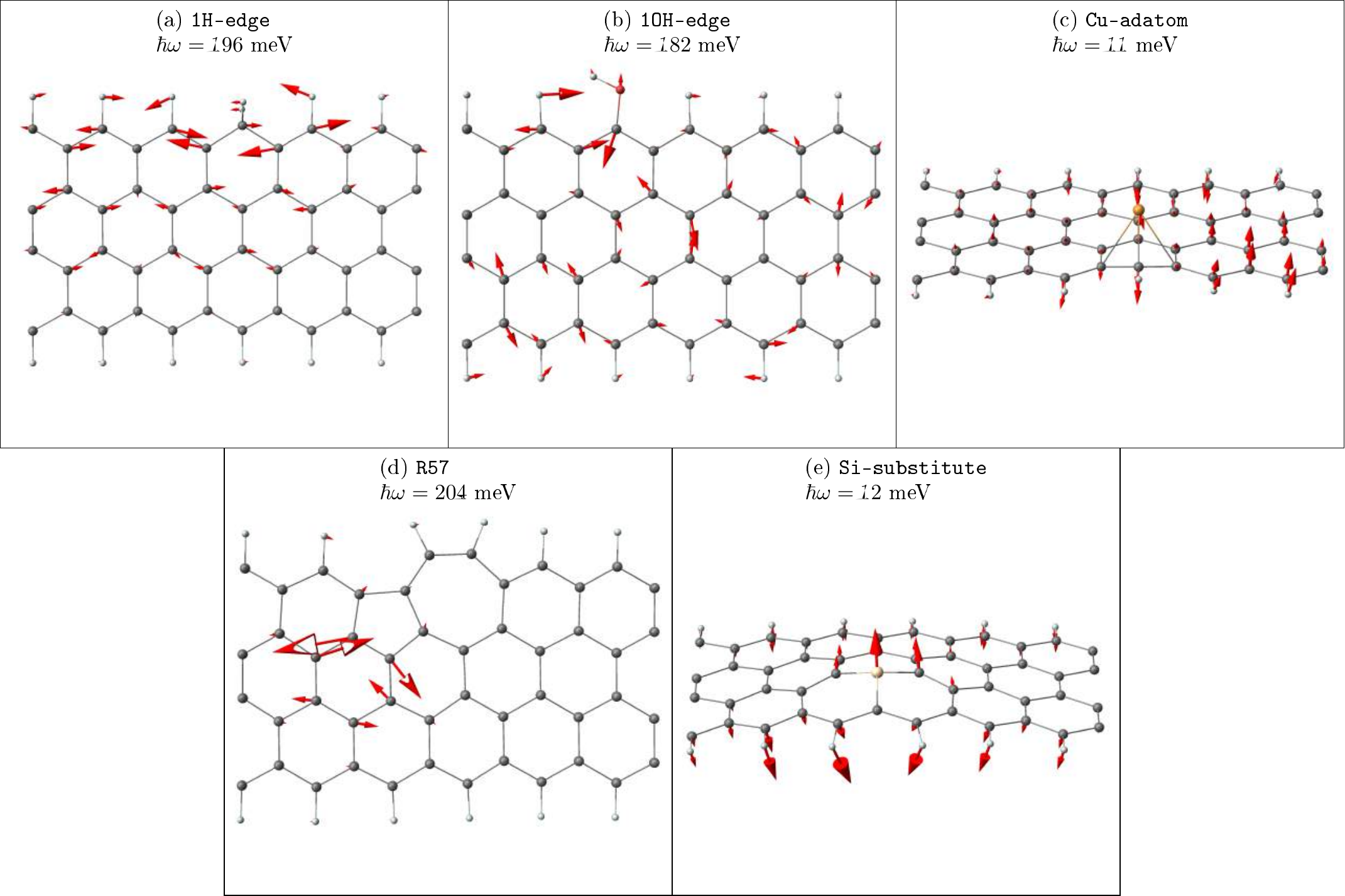}
 \caption{\label{fig:ZigModes} (Color online)  Visualization of the most contributing modes to the IETS signals indicated by vertical lines in \Figref{fig:IETSdefectZGNR}(j-o).
(a) Edge mode at the edge with the extra hydrogen in \texttt{1H-edge}.  (b) Mode contributing to the asymmetric signal in \Figref{fig:IETSdefectZGNR} (m) for the \texttt{1OH-edge}. (c) Out-of-plane mode for \texttt{Cu-adatom}. (d) Localized mode for \texttt{R57}. (e) Out o plane mode for \texttt{Si-substitute}.
}
\end{figure*}

Substituting a carbon with a silicon atom leads to an out-of-plane buckling, see \Figref{fig:StructDefectZGNR}(i). However, both silicon and carbon have an $s^2p^2$ electronic structure, and the electron transmission is basically similar to the pristine. On the other hand, the buckling give rise to low energy peaks in the IETS signal originating from the e-ph coupling to the out-of-plane modes [\Figref{fig:ZigModes}(e)]. Gating close to the band edge of the conduction band gives rise to different sign of the signals at low and high vibrational energies, as seen from the top curve in \Figref{fig:IETSdefectZGNR}(o).

\section{Conclusions}

In summary, we have investigated IETS signals in symmetrically contacted armchair and zigzag graphene nanoribbons, considering both pristine as well as a selection of defected configurations under varying charge carrier conditions.
For the clean AGNR inelastic tunneling gives rise to two distinct peaks in the IETS spectrum at $169$ mV and $196$ mV corresponding to the D-and G-modes of Raman spectroscopy, respectively.  By connecting the IETS signals to the phonon band structure, we have clarified how only a single band contributes to the "G-mode" while three bands contribute to the broader "D-mode". 
Concerning defects in AGNRs we have shown how some leave IETS unchanged while others give clear signals.
For instance, adding an extra hydrogen atom to a single edge side gives a clear signal for some gate values. This signal can be removed by adding another hydrogen atom to the neighboring edge side because the sub-lattice symmetry is restored. 
Further, exchanging a single hydrogen atom with a fluorine atom in the passivation does not result in any change in both the elastic and inelastic tunneling.
However removing 8 hydrogen atoms leaving part of the edge on each side without passivation, gives a clear robust signal throughout the investigated gate values.  The signal, due to the vibration of the carbon dimers at the edge, has an energy around $245$ meV making it easy to detect since it is outside the vibrational spectrum of the pristine ribbon.   
Breaking of one or two C-C bonds turns out to interchange the relative intensity of the "G"- and "D"-peaks. 
Breaking 4 C-C bonds gives rise to signals caused by the defect tilted out of plane.
Lifting the symmetry of the $\pi$-electrons by adding a Cu-adatom allows the out-of-plane modes to contribute.

For the ZGNR we find relatively broader IETS signals especially in the absense of a large gate voltage ($V_G\approx 0$ V).
Importantly, this  is a consequence of the breaking of the axial mirror symmetry in the ribbon due to  the presence of spin-polarized edges. 
Thus, by comparing to spin-degenerate calculations, we suggest that IETS can give an indirect proof of spin-polarization in zigzag ribbons.     
On the other hand, the broader IETS features may make it difficult to identify the different defect signals reported in this paper.

The presence of a \texttt{R57}-reconstruction also broadens the IETS by breaking both the electronic and vibrational symmetry. 
Substituting a carbon atom with a silicon atom makes the ribbon buckle, breaking the planar symmetry, allowing the out-of-plane modes to contribute to the IETS.
This suggests that IETS in principle could be used to gain information of the curvature of GNRs and other graphene-based structures. 

Finally, as an outlook we note that here we presented calculations on long, symmetrically contacted systems where there is a significant overlap with both metallic electrodes. It would be interesting to extend such a study also to the  asymmetric situation where a point tunnel contact is made to one end resembling, say, the coupling a STM tip.\cite{VanderLit2013f}


\begin{thebibliography}{54}
	\expandafter\ifx\csname natexlab\endcsname\relax\def\natexlab#1{#1}\fi
	\expandafter\ifx\csname bibnamefont\endcsname\relax
	\def\bibnamefont#1{#1}\fi
	\expandafter\ifx\csname bibfnamefont\endcsname\relax
	\def\bibfnamefont#1{#1}\fi
	\expandafter\ifx\csname citenamefont\endcsname\relax
	\def\citenamefont#1{#1}\fi
	\expandafter\ifx\csname url\endcsname\relax
	\def\url#1{\texttt{#1}}\fi
	\expandafter\ifx\csname urlprefix\endcsname\relax\def\urlprefix{URL }\fi
	\providecommand{\bibinfo}[2]{#2}
	\providecommand{\eprint}[2][]{\url{#2}}
	
	\bibitem[{\citenamefont{Raza}(2011)}]{Raza2011}
	\bibinfo{author}{\bibfnamefont{H.}~\bibnamefont{Raza}},
	\emph{\bibinfo{title}{{Graphene nanoelectronics : metrology, synthesis,
				properties and applications}}} (\bibinfo{publisher}{Springer},
	\bibinfo{year}{2011}).
	
	\bibitem[{\citenamefont{Foa~Torres et~al.}(2014)\citenamefont{Foa~Torres,
			Roche, and Charlier}}]{FoaTorres2014}
	\bibinfo{author}{\bibfnamefont{L.~E.~F.} \bibnamefont{Foa~Torres}},
	\bibinfo{author}{\bibfnamefont{S.}~\bibnamefont{Roche}}, \bibnamefont{and}
	\bibinfo{author}{\bibfnamefont{J.-C.} \bibnamefont{Charlier}},
	\emph{\bibinfo{title}{Introduction to Graphene-Based Nanomaterials: From
			Electronic Structure to Quantum Transport}} (\bibinfo{publisher}{Cambridge
		University Press}, \bibinfo{year}{2014}).
	
	\bibitem[{\citenamefont{Dutta and Pati}(2010)}]{Dutta2010}
	\bibinfo{author}{\bibfnamefont{S.}~\bibnamefont{Dutta}} \bibnamefont{and}
	\bibinfo{author}{\bibfnamefont{S.~K.} \bibnamefont{Pati}},
	\bibinfo{journal}{J. Mater. Chem.} \textbf{\bibinfo{volume}{20}},
	\bibinfo{pages}{8207} (\bibinfo{year}{2010}).
	
	\bibitem[{\citenamefont{Pedersen et~al.}(2008)\citenamefont{Pedersen, Flindt,
			Pedersen, Mortensen, Jauho, and Pedersen}}]{pedersen_graphene_2008}
	\bibinfo{author}{\bibfnamefont{T.~G.} \bibnamefont{Pedersen}},
	\bibinfo{author}{\bibfnamefont{C.}~\bibnamefont{Flindt}},
	\bibinfo{author}{\bibfnamefont{J.}~\bibnamefont{Pedersen}},
	\bibinfo{author}{\bibfnamefont{N.~A.} \bibnamefont{Mortensen}},
	\bibinfo{author}{\bibfnamefont{A.}~\bibnamefont{Jauho}}, \bibnamefont{and}
	\bibinfo{author}{\bibfnamefont{K.}~\bibnamefont{Pedersen}},
	\bibinfo{journal}{Phys.~Rev.~Lett.} \textbf{\bibinfo{volume}{100}},
	\bibinfo{pages}{136804} (\bibinfo{year}{2008}).
	
	\bibitem[{\citenamefont{Bai et~al.}(2010)\citenamefont{Bai, Zhong, Jiang,
			Huang, and Duan}}]{Bai2010}
	\bibinfo{author}{\bibfnamefont{J.}~\bibnamefont{Bai}},
	\bibinfo{author}{\bibfnamefont{X.}~\bibnamefont{Zhong}},
	\bibinfo{author}{\bibfnamefont{S.}~\bibnamefont{Jiang}},
	\bibinfo{author}{\bibfnamefont{Y.}~\bibnamefont{Huang}}, \bibnamefont{and}
	\bibinfo{author}{\bibfnamefont{X.}~\bibnamefont{Duan}},
	\bibinfo{journal}{Nat.~ Nanotechnol.} \textbf{\bibinfo{volume}{5}},
	\bibinfo{pages}{190} (\bibinfo{year}{2010}).
	
	\bibitem[{\citenamefont{Wagner et~al.}(2013{\natexlab{a}})\citenamefont{Wagner,
			Ewels, Adjizian, Magaud, Pochet, Roche, Lopez-Bezanilla, Ivanovskaya, Yaya,
			Rayson et~al.}}]{Wagner2013}
	\bibinfo{author}{\bibfnamefont{P.}~\bibnamefont{Wagner}},
	\bibinfo{author}{\bibfnamefont{C.~P.} \bibnamefont{Ewels}},
	\bibinfo{author}{\bibfnamefont{J.-J.} \bibnamefont{Adjizian}},
	\bibinfo{author}{\bibfnamefont{L.}~\bibnamefont{Magaud}},
	\bibinfo{author}{\bibfnamefont{P.}~\bibnamefont{Pochet}},
	\bibinfo{author}{\bibfnamefont{S.}~\bibnamefont{Roche}},
	\bibinfo{author}{\bibfnamefont{A.}~\bibnamefont{Lopez-Bezanilla}},
	\bibinfo{author}{\bibfnamefont{V.~V.} \bibnamefont{Ivanovskaya}},
	\bibinfo{author}{\bibfnamefont{A.}~\bibnamefont{Yaya}},
	\bibinfo{author}{\bibfnamefont{M.}~\bibnamefont{Rayson}},
	\bibnamefont{et~al.}, \bibinfo{journal}{J. Phys. Chem. C}
	\textbf{\bibinfo{volume}{117}}, \bibinfo{pages}{26790}
	(\bibinfo{year}{2013}{\natexlab{a}}).
	
	\bibitem[{\citenamefont{Ferrari and Basko}(2013)}]{Ferrari2013}
	\bibinfo{author}{\bibfnamefont{A.~C.} \bibnamefont{Ferrari}} \bibnamefont{and}
	\bibinfo{author}{\bibfnamefont{D.~M.} \bibnamefont{Basko}},
	\bibinfo{journal}{Nat.~Nanotechnol.} \textbf{\bibinfo{volume}{8}},
	\bibinfo{pages}{235} (\bibinfo{year}{2013}).
	
	\bibitem[{\citenamefont{Shiotari et~al.}(2014)\citenamefont{Shiotari, Kumagai,
			and Wolf}}]{Shiotari2014}
	\bibinfo{author}{\bibfnamefont{A.}~\bibnamefont{Shiotari}},
	\bibinfo{author}{\bibfnamefont{T.}~\bibnamefont{Kumagai}}, \bibnamefont{and}
	\bibinfo{author}{\bibfnamefont{M.}~\bibnamefont{Wolf}}, \bibinfo{journal}{The
		Journal of Physical Chemistry C} \textbf{\bibinfo{volume}{118}},
	\bibinfo{pages}{11806} (\bibinfo{year}{2014}).
	
	\bibitem[{\citenamefont{Han et~al.}(2007)\citenamefont{Han, \"{O}zyilmaz,
			Zhang, and Kim}}]{Han2007}
	\bibinfo{author}{\bibfnamefont{M.~Y.} \bibnamefont{Han}},
	\bibinfo{author}{\bibfnamefont{B.}~\bibnamefont{\"{O}zyilmaz}},
	\bibinfo{author}{\bibfnamefont{Y.}~\bibnamefont{Zhang}}, \bibnamefont{and}
	\bibinfo{author}{\bibfnamefont{P.}~\bibnamefont{Kim}},
	\bibinfo{journal}{Phys. Rev. Lett.} \textbf{\bibinfo{volume}{98}},
	\bibinfo{pages}{206805} (\bibinfo{year}{2007}).
	
	\bibitem[{\citenamefont{Li et~al.}(2008)\citenamefont{Li, Wang, Zhang, Lee, and
			Dai}}]{Li2008a}
	\bibinfo{author}{\bibfnamefont{X.}~\bibnamefont{Li}},
	\bibinfo{author}{\bibfnamefont{X.}~\bibnamefont{Wang}},
	\bibinfo{author}{\bibfnamefont{L.}~\bibnamefont{Zhang}},
	\bibinfo{author}{\bibfnamefont{S.}~\bibnamefont{Lee}}, \bibnamefont{and}
	\bibinfo{author}{\bibfnamefont{H.}~\bibnamefont{Dai}},
	\bibinfo{journal}{Science} \textbf{\bibinfo{volume}{319}},
	\bibinfo{pages}{1229} (\bibinfo{year}{2008}).
	
	\bibitem[{\citenamefont{Wu et~al.}(2010)\citenamefont{Wu, Ren, Gao, Liu, Zhao,
			and Cheng}}]{Wu2010}
	\bibinfo{author}{\bibfnamefont{Z.-S.} \bibnamefont{Wu}},
	\bibinfo{author}{\bibfnamefont{W.}~\bibnamefont{Ren}},
	\bibinfo{author}{\bibfnamefont{L.}~\bibnamefont{Gao}},
	\bibinfo{author}{\bibfnamefont{B.}~\bibnamefont{Liu}},
	\bibinfo{author}{\bibfnamefont{J.}~\bibnamefont{Zhao}}, \bibnamefont{and}
	\bibinfo{author}{\bibfnamefont{H.-M.} \bibnamefont{Cheng}},
	\bibinfo{journal}{Nano Res.} \textbf{\bibinfo{volume}{3}},
	\bibinfo{pages}{16} (\bibinfo{year}{2010}).
	
	\bibitem[{\citenamefont{Baringhaus et~al.}(2014)\citenamefont{Baringhaus, Ruan,
			Edler, Tejeda, Sicot, Taleb-Ibrahimi, Li, Jiang, Conrad, Berger
			et~al.}}]{Baringhaus2014}
	\bibinfo{author}{\bibfnamefont{J.}~\bibnamefont{Baringhaus}},
	\bibinfo{author}{\bibfnamefont{M.}~\bibnamefont{Ruan}},
	\bibinfo{author}{\bibfnamefont{F.}~\bibnamefont{Edler}},
	\bibinfo{author}{\bibfnamefont{A.}~\bibnamefont{Tejeda}},
	\bibinfo{author}{\bibfnamefont{M.}~\bibnamefont{Sicot}},
	\bibinfo{author}{\bibfnamefont{A.}~\bibnamefont{Taleb-Ibrahimi}},
	\bibinfo{author}{\bibfnamefont{A.-P.} \bibnamefont{Li}},
	\bibinfo{author}{\bibfnamefont{Z.}~\bibnamefont{Jiang}},
	\bibinfo{author}{\bibfnamefont{E.~H.} \bibnamefont{Conrad}},
	\bibinfo{author}{\bibfnamefont{C.}~\bibnamefont{Berger}},
	\bibnamefont{et~al.}, \bibinfo{journal}{Nature}
	\textbf{\bibinfo{volume}{506}}, \bibinfo{pages}{349} (\bibinfo{year}{2014}).
	
	\bibitem[{\citenamefont{Kosynkin et~al.}(2009)\citenamefont{Kosynkin,
			Higginbotham, Sinitskii, Lomeda, Dimiev, Price, and Tour}}]{Kosynkin2009}
	\bibinfo{author}{\bibfnamefont{D.~V.} \bibnamefont{Kosynkin}},
	\bibinfo{author}{\bibfnamefont{A.~L.} \bibnamefont{Higginbotham}},
	\bibinfo{author}{\bibfnamefont{A.}~\bibnamefont{Sinitskii}},
	\bibinfo{author}{\bibfnamefont{J.~R.} \bibnamefont{Lomeda}},
	\bibinfo{author}{\bibfnamefont{A.}~\bibnamefont{Dimiev}},
	\bibinfo{author}{\bibfnamefont{B.~K.} \bibnamefont{Price}}, \bibnamefont{and}
	\bibinfo{author}{\bibfnamefont{J.~M.} \bibnamefont{Tour}},
	\bibinfo{journal}{Nature} \textbf{\bibinfo{volume}{458}},
	\bibinfo{pages}{872} (\bibinfo{year}{2009}).
	
	\bibitem[{\citenamefont{Cai et~al.}(2010)\citenamefont{Cai, Ruffieux, Jaafar,
			Bieri, Braun, Blankenburg, Muoth, Seitsonen, Saleh, Feng et~al.}}]{Cai2010}
	\bibinfo{author}{\bibfnamefont{J.}~\bibnamefont{Cai}},
	\bibinfo{author}{\bibfnamefont{P.}~\bibnamefont{Ruffieux}},
	\bibinfo{author}{\bibfnamefont{R.}~\bibnamefont{Jaafar}},
	\bibinfo{author}{\bibfnamefont{M.}~\bibnamefont{Bieri}},
	\bibinfo{author}{\bibfnamefont{T.}~\bibnamefont{Braun}},
	\bibinfo{author}{\bibfnamefont{S.}~\bibnamefont{Blankenburg}},
	\bibinfo{author}{\bibfnamefont{M.}~\bibnamefont{Muoth}},
	\bibinfo{author}{\bibfnamefont{A.~P.} \bibnamefont{Seitsonen}},
	\bibinfo{author}{\bibfnamefont{M.}~\bibnamefont{Saleh}},
	\bibinfo{author}{\bibfnamefont{X.}~\bibnamefont{Feng}}, \bibnamefont{et~al.},
	\bibinfo{journal}{Nature} \textbf{\bibinfo{volume}{466}},
	\bibinfo{pages}{470} (\bibinfo{year}{2010}).
	
	\bibitem[{\citenamefont{Blankenburg et~al.}(2012)\citenamefont{Blankenburg,
			Cai, Ruffieux, Jaafar, Passerone, Feng, M\"{u}llen, Fasel, and
			Pignedoli}}]{Blankenburg2012}
	\bibinfo{author}{\bibfnamefont{S.}~\bibnamefont{Blankenburg}},
	\bibinfo{author}{\bibfnamefont{J.}~\bibnamefont{Cai}},
	\bibinfo{author}{\bibfnamefont{P.}~\bibnamefont{Ruffieux}},
	\bibinfo{author}{\bibfnamefont{R.}~\bibnamefont{Jaafar}},
	\bibinfo{author}{\bibfnamefont{D.}~\bibnamefont{Passerone}},
	\bibinfo{author}{\bibfnamefont{X.}~\bibnamefont{Feng}},
	\bibinfo{author}{\bibfnamefont{K.}~\bibnamefont{M\"{u}llen}},
	\bibinfo{author}{\bibfnamefont{R.}~\bibnamefont{Fasel}}, \bibnamefont{and}
	\bibinfo{author}{\bibfnamefont{C.~a.} \bibnamefont{Pignedoli}},
	\bibinfo{journal}{ACS Nano} \textbf{\bibinfo{volume}{6}},
	\bibinfo{pages}{2020} (\bibinfo{year}{2012}).
	
	\bibitem[{\citenamefont{Han et~al.}(2014)\citenamefont{Han, Akagi, {Federici
				Canova}, Mutoh, Shiraki, Iwaya, Weiss, Asao, and Hitosugi}}]{Han2014}
	\bibinfo{author}{\bibfnamefont{P.}~\bibnamefont{Han}},
	\bibinfo{author}{\bibfnamefont{K.}~\bibnamefont{Akagi}},
	\bibinfo{author}{\bibfnamefont{F.}~\bibnamefont{{Federici Canova}}},
	\bibinfo{author}{\bibfnamefont{H.}~\bibnamefont{Mutoh}},
	\bibinfo{author}{\bibfnamefont{S.}~\bibnamefont{Shiraki}},
	\bibinfo{author}{\bibfnamefont{K.}~\bibnamefont{Iwaya}},
	\bibinfo{author}{\bibfnamefont{P.~S.} \bibnamefont{Weiss}},
	\bibinfo{author}{\bibfnamefont{N.}~\bibnamefont{Asao}}, \bibnamefont{and}
	\bibinfo{author}{\bibfnamefont{T.}~\bibnamefont{Hitosugi}},
	\bibinfo{journal}{ACS Nano} \textbf{\bibinfo{volume}{8}},
	\bibinfo{pages}{9181} (\bibinfo{year}{2014}).
	
	\bibitem[{\citenamefont{Cai et~al.}(2014)\citenamefont{Cai, Pignedoli, Talirz,
			Ruffieux, S\"ode, Liang, Meunier, Berger, Li, Feng et~al.}}]{Cai2014}
	\bibinfo{author}{\bibfnamefont{J.}~\bibnamefont{Cai}},
	\bibinfo{author}{\bibfnamefont{C.~A.} \bibnamefont{Pignedoli}},
	\bibinfo{author}{\bibfnamefont{L.}~\bibnamefont{Talirz}},
	\bibinfo{author}{\bibfnamefont{P.}~\bibnamefont{Ruffieux}},
	\bibinfo{author}{\bibfnamefont{H.}~\bibnamefont{S\"ode}},
	\bibinfo{author}{\bibfnamefont{L.}~\bibnamefont{Liang}},
	\bibinfo{author}{\bibfnamefont{V.}~\bibnamefont{Meunier}},
	\bibinfo{author}{\bibfnamefont{R.}~\bibnamefont{Berger}},
	\bibinfo{author}{\bibfnamefont{R.}~\bibnamefont{Li}},
	\bibinfo{author}{\bibfnamefont{X.}~\bibnamefont{Feng}}, \bibnamefont{et~al.},
	\bibinfo{journal}{Nat.~Nanotechnol.} \textbf{\bibinfo{volume}{9}},
	\bibinfo{pages}{896} (\bibinfo{year}{2014}).
	
	\bibitem[{\citenamefont{Bronner et~al.}(2012)\citenamefont{Bronner, Leyssner,
			Stremlau, Utecht, Saalfrank, Klamroth, and Tegeder}}]{Bronner2012}
	\bibinfo{author}{\bibfnamefont{C.}~\bibnamefont{Bronner}},
	\bibinfo{author}{\bibfnamefont{F.}~\bibnamefont{Leyssner}},
	\bibinfo{author}{\bibfnamefont{S.}~\bibnamefont{Stremlau}},
	\bibinfo{author}{\bibfnamefont{M.}~\bibnamefont{Utecht}},
	\bibinfo{author}{\bibfnamefont{P.}~\bibnamefont{Saalfrank}},
	\bibinfo{author}{\bibfnamefont{T.}~\bibnamefont{Klamroth}}, \bibnamefont{and}
	\bibinfo{author}{\bibfnamefont{P.}~\bibnamefont{Tegeder}},
	\bibinfo{journal}{Phys. Rev. B} \textbf{\bibinfo{volume}{86}},
	\bibinfo{pages}{085444} (\bibinfo{year}{2012}).
	
	\bibitem[{\citenamefont{Ruffieux et~al.}(2012)\citenamefont{Ruffieux, Cai,
			Plumb, Patthey, Prezzi, Ferretti, Molinari, Feng, M\"{u}llen, Pignedoli
			et~al.}}]{Ruffieux2012}
	\bibinfo{author}{\bibfnamefont{P.}~\bibnamefont{Ruffieux}},
	\bibinfo{author}{\bibfnamefont{J.}~\bibnamefont{Cai}},
	\bibinfo{author}{\bibfnamefont{N.~C.} \bibnamefont{Plumb}},
	\bibinfo{author}{\bibfnamefont{L.}~\bibnamefont{Patthey}},
	\bibinfo{author}{\bibfnamefont{D.}~\bibnamefont{Prezzi}},
	\bibinfo{author}{\bibfnamefont{A.}~\bibnamefont{Ferretti}},
	\bibinfo{author}{\bibfnamefont{E.}~\bibnamefont{Molinari}},
	\bibinfo{author}{\bibfnamefont{X.}~\bibnamefont{Feng}},
	\bibinfo{author}{\bibfnamefont{K.}~\bibnamefont{M\"{u}llen}},
	\bibinfo{author}{\bibfnamefont{C.~a.} \bibnamefont{Pignedoli}},
	\bibnamefont{et~al.}, \bibinfo{journal}{ACS Nano}
	\textbf{\bibinfo{volume}{6}}, \bibinfo{pages}{6930} (\bibinfo{year}{2012}).
	
	\bibitem[{\citenamefont{van~der Lit et~al.}(2013)\citenamefont{van~der Lit,
			Boneschanscher, Vanmaekelbergh, Ij\"{a}s, Uppstu, Ervasti, Harju, Liljeroth,
			and Swart}}]{VanderLit2013f}
	\bibinfo{author}{\bibfnamefont{J.}~\bibnamefont{van~der Lit}},
	\bibinfo{author}{\bibfnamefont{M.~P.} \bibnamefont{Boneschanscher}},
	\bibinfo{author}{\bibfnamefont{D.}~\bibnamefont{Vanmaekelbergh}},
	\bibinfo{author}{\bibfnamefont{M.}~\bibnamefont{Ij\"{a}s}},
	\bibinfo{author}{\bibfnamefont{A.}~\bibnamefont{Uppstu}},
	\bibinfo{author}{\bibfnamefont{M.}~\bibnamefont{Ervasti}},
	\bibinfo{author}{\bibfnamefont{A.}~\bibnamefont{Harju}},
	\bibinfo{author}{\bibfnamefont{P.}~\bibnamefont{Liljeroth}},
	\bibnamefont{and} \bibinfo{author}{\bibfnamefont{I.}~\bibnamefont{Swart}},
	\bibinfo{journal}{Nat. Commun.} \textbf{\bibinfo{volume}{4}},
	\bibinfo{pages}{2023} (\bibinfo{year}{2013}).
	
	\bibitem[{\citenamefont{Koch et~al.}(2012)\citenamefont{Koch, Ample, Joachim,
			and Grill}}]{Koch2012}
	\bibinfo{author}{\bibfnamefont{M.}~\bibnamefont{Koch}},
	\bibinfo{author}{\bibfnamefont{F.}~\bibnamefont{Ample}},
	\bibinfo{author}{\bibfnamefont{C.}~\bibnamefont{Joachim}}, \bibnamefont{and}
	\bibinfo{author}{\bibfnamefont{L.}~\bibnamefont{Grill}},
	\bibinfo{journal}{Nat. Nanotechnol.} \textbf{\bibinfo{volume}{7}},
	\bibinfo{pages}{713} (\bibinfo{year}{2012}).
	
	\bibitem[{\citenamefont{Wassmann et~al.}(2008)\citenamefont{Wassmann,
			Seitsonen, Saitta, Lazzeri, and Mauri}}]{Wassmann2008}
	\bibinfo{author}{\bibfnamefont{T.}~\bibnamefont{Wassmann}},
	\bibinfo{author}{\bibfnamefont{A.~P.} \bibnamefont{Seitsonen}},
	\bibinfo{author}{\bibfnamefont{A.~M.} \bibnamefont{Saitta}},
	\bibinfo{author}{\bibfnamefont{M.}~\bibnamefont{Lazzeri}}, \bibnamefont{and}
	\bibinfo{author}{\bibfnamefont{F.}~\bibnamefont{Mauri}},
	\bibinfo{journal}{Phys. Rev. Lett.} \textbf{\bibinfo{volume}{101}},
	\bibinfo{pages}{096402} (\bibinfo{year}{2008}).
	
	\bibitem[{\citenamefont{Li et~al.}(2010)\citenamefont{Li, Li, Zhou, Liu, Wu,
			Gu, Ihm, and Duan}}]{Li2010c}
	\bibinfo{author}{\bibfnamefont{J.}~\bibnamefont{Li}},
	\bibinfo{author}{\bibfnamefont{Z.}~\bibnamefont{Li}},
	\bibinfo{author}{\bibfnamefont{G.}~\bibnamefont{Zhou}},
	\bibinfo{author}{\bibfnamefont{Z.}~\bibnamefont{Liu}},
	\bibinfo{author}{\bibfnamefont{J.}~\bibnamefont{Wu}},
	\bibinfo{author}{\bibfnamefont{B.-L.} \bibnamefont{Gu}},
	\bibinfo{author}{\bibfnamefont{J.}~\bibnamefont{Ihm}}, \bibnamefont{and}
	\bibinfo{author}{\bibfnamefont{W.}~\bibnamefont{Duan}},
	\bibinfo{journal}{Phys. Rev. B} \textbf{\bibinfo{volume}{82}},
	\bibinfo{pages}{115410} (\bibinfo{year}{2010}).
	
	\bibitem[{\citenamefont{Wagner et~al.}(2013{\natexlab{b}})\citenamefont{Wagner,
			Ivanovskaya, Melle-Franco, Humbert, Adjizian, Briddon, and
			Ewels}}]{Wagner2013a}
	\bibinfo{author}{\bibfnamefont{P.}~\bibnamefont{Wagner}},
	\bibinfo{author}{\bibfnamefont{V.~V.} \bibnamefont{Ivanovskaya}},
	\bibinfo{author}{\bibfnamefont{M.}~\bibnamefont{Melle-Franco}},
	\bibinfo{author}{\bibfnamefont{B.}~\bibnamefont{Humbert}},
	\bibinfo{author}{\bibfnamefont{J.-J.} \bibnamefont{Adjizian}},
	\bibinfo{author}{\bibfnamefont{P.~R.} \bibnamefont{Briddon}},
	\bibnamefont{and} \bibinfo{author}{\bibfnamefont{C.~P.} \bibnamefont{Ewels}},
	\bibinfo{journal}{Phys. Rev. B} \textbf{\bibinfo{volume}{88}},
	\bibinfo{pages}{094106} (\bibinfo{year}{2013}{\natexlab{b}}).
	
	\bibitem[{\citenamefont{Hod et~al.}(2007)\citenamefont{Hod, Barone, Peralta,
			and Scuseria}}]{Hod2007}
	\bibinfo{author}{\bibfnamefont{O.}~\bibnamefont{Hod}},
	\bibinfo{author}{\bibfnamefont{V.}~\bibnamefont{Barone}},
	\bibinfo{author}{\bibfnamefont{J.~E.} \bibnamefont{Peralta}},
	\bibnamefont{and} \bibinfo{author}{\bibfnamefont{G.~E.}
		\bibnamefont{Scuseria}}, \bibinfo{journal}{Nano Lett.}
	\textbf{\bibinfo{volume}{7}}, \bibinfo{pages}{2295} (\bibinfo{year}{2007}).
	
	\bibitem[{\citenamefont{Gunlycke et~al.}(2007)\citenamefont{Gunlycke, Li,
			Mintmire, and White}}]{Gunlycke2007}
	\bibinfo{author}{\bibfnamefont{D.}~\bibnamefont{Gunlycke}},
	\bibinfo{author}{\bibfnamefont{J.}~\bibnamefont{Li}},
	\bibinfo{author}{\bibfnamefont{J.~W.} \bibnamefont{Mintmire}},
	\bibnamefont{and} \bibinfo{author}{\bibfnamefont{C.~T.} \bibnamefont{White}},
	\bibinfo{journal}{Appl. Phys. Lett.} \textbf{\bibinfo{volume}{91}},
	\bibinfo{pages}{112108} (\bibinfo{year}{2007}).
	
	\bibitem[{\citenamefont{Zhang et~al.}(2012)\citenamefont{Zhang, He, Xue, Zhang,
			Sun, and Zhong}}]{Zhang2012a}
	\bibinfo{author}{\bibfnamefont{C.}~\bibnamefont{Zhang}},
	\bibinfo{author}{\bibfnamefont{C.}~\bibnamefont{He}},
	\bibinfo{author}{\bibfnamefont{L.}~\bibnamefont{Xue}},
	\bibinfo{author}{\bibfnamefont{K.}~\bibnamefont{Zhang}},
	\bibinfo{author}{\bibfnamefont{L.}~\bibnamefont{Sun}}, \bibnamefont{and}
	\bibinfo{author}{\bibfnamefont{J.}~\bibnamefont{Zhong}},
	\bibinfo{journal}{Org. Electron.} \textbf{\bibinfo{volume}{13}},
	\bibinfo{pages}{2494} (\bibinfo{year}{2012}).
	
	\bibitem[{\citenamefont{Al-Aqtash et~al.}(2013)\citenamefont{Al-Aqtash, Li,
			Wang, Mei, and Sabirianov}}]{Al-Aqtash2013}
	\bibinfo{author}{\bibfnamefont{N.}~\bibnamefont{Al-Aqtash}},
	\bibinfo{author}{\bibfnamefont{H.}~\bibnamefont{Li}},
	\bibinfo{author}{\bibfnamefont{L.}~\bibnamefont{Wang}},
	\bibinfo{author}{\bibfnamefont{W.-N.} \bibnamefont{Mei}}, \bibnamefont{and}
	\bibinfo{author}{\bibfnamefont{R.}~\bibnamefont{Sabirianov}},
	\bibinfo{journal}{Carbon} \textbf{\bibinfo{volume}{51}}, \bibinfo{pages}{102}
	(\bibinfo{year}{2013}).
	
	\bibitem[{\citenamefont{Yamamoto et~al.}(2004)\citenamefont{Yamamoto, Watanabe,
			and Mii}}]{Yamamoto2004}
	\bibinfo{author}{\bibfnamefont{T.}~\bibnamefont{Yamamoto}},
	\bibinfo{author}{\bibfnamefont{K.}~\bibnamefont{Watanabe}}, \bibnamefont{and}
	\bibinfo{author}{\bibfnamefont{K.}~\bibnamefont{Mii}},
	\bibinfo{journal}{Phys.~Rev.~B} \textbf{\bibinfo{volume}{70}},
	\bibinfo{pages}{245402} (\bibinfo{year}{2004}).
	
	\bibitem[{\citenamefont{Vandescuren et~al.}(2008)\citenamefont{Vandescuren,
			Hermet, Meunier, Henrard, and Lambin}}]{Vandescuren2008}
	\bibinfo{author}{\bibfnamefont{M.}~\bibnamefont{Vandescuren}},
	\bibinfo{author}{\bibfnamefont{P.}~\bibnamefont{Hermet}},
	\bibinfo{author}{\bibfnamefont{V.}~\bibnamefont{Meunier}},
	\bibinfo{author}{\bibfnamefont{L.}~\bibnamefont{Henrard}}, \bibnamefont{and}
	\bibinfo{author}{\bibfnamefont{P.}~\bibnamefont{Lambin}},
	\bibinfo{journal}{Phys. Rev. B} \textbf{\bibinfo{volume}{78}},
	\bibinfo{pages}{195401} (\bibinfo{year}{2008}).
	
	\bibitem[{\citenamefont{Gillen et~al.}(2009)\citenamefont{Gillen, Mohr,
			Maultzsch, and Thomsen}}]{Gillen2009}
	\bibinfo{author}{\bibfnamefont{R.}~\bibnamefont{Gillen}},
	\bibinfo{author}{\bibfnamefont{M.}~\bibnamefont{Mohr}},
	\bibinfo{author}{\bibfnamefont{J.}~\bibnamefont{Maultzsch}},
	\bibnamefont{and} \bibinfo{author}{\bibfnamefont{C.}~\bibnamefont{Thomsen}},
	\bibinfo{journal}{Phys. Status Solidi} \textbf{\bibinfo{volume}{246}},
	\bibinfo{pages}{2577} (\bibinfo{year}{2009}).
	
	\bibitem[{\citenamefont{Zhou and Dong}(2007)}]{Zhou2007}
	\bibinfo{author}{\bibfnamefont{J.}~\bibnamefont{Zhou}} \bibnamefont{and}
	\bibinfo{author}{\bibfnamefont{J.}~\bibnamefont{Dong}},
	\bibinfo{journal}{Appl. Phys. Lett.} \textbf{\bibinfo{volume}{91}},
	\bibinfo{pages}{173108} (\bibinfo{year}{2007}).
	
	\bibitem[{\citenamefont{Saito et~al.}(2010)\citenamefont{Saito, Furukawa,
			Dresselhaus, and Dresselhaus}}]{Saito2010a}
	\bibinfo{author}{\bibfnamefont{R.}~\bibnamefont{Saito}},
	\bibinfo{author}{\bibfnamefont{M.}~\bibnamefont{Furukawa}},
	\bibinfo{author}{\bibfnamefont{G.}~\bibnamefont{Dresselhaus}},
	\bibnamefont{and} \bibinfo{author}{\bibfnamefont{M.~S.}
		\bibnamefont{Dresselhaus}}, \bibinfo{journal}{J. Phys. Condens. Matter}
	\textbf{\bibinfo{volume}{22}}, \bibinfo{pages}{334203}
	(\bibinfo{year}{2010}).
	
	\bibitem[{\citenamefont{Huang et~al.}(2012)\citenamefont{Huang, Wei, Sun, Wong,
			Feng, Neto, and Wee}}]{Huang2012b}
	\bibinfo{author}{\bibfnamefont{H.}~\bibnamefont{Huang}},
	\bibinfo{author}{\bibfnamefont{D.}~\bibnamefont{Wei}},
	\bibinfo{author}{\bibfnamefont{J.}~\bibnamefont{Sun}},
	\bibinfo{author}{\bibfnamefont{S.~L.} \bibnamefont{Wong}},
	\bibinfo{author}{\bibfnamefont{Y.~P.} \bibnamefont{Feng}},
	\bibinfo{author}{\bibfnamefont{a.~H.~C.} \bibnamefont{Neto}},
	\bibnamefont{and} \bibinfo{author}{\bibfnamefont{A.~T.~S.}
		\bibnamefont{Wee}}, \bibinfo{journal}{Sci. Rep.}
	\textbf{\bibinfo{volume}{2}}, \bibinfo{pages}{983} (\bibinfo{year}{2012}).
	
	\bibitem[{\citenamefont{Ij\"{a}s et~al.}(2013)\citenamefont{Ij\"{a}s, Ervasti,
			Uppstu, Liljeroth, van~der Lit, Swart, and Harju}}]{Ijas2013}
	\bibinfo{author}{\bibfnamefont{M.}~\bibnamefont{Ij\"{a}s}},
	\bibinfo{author}{\bibfnamefont{M.}~\bibnamefont{Ervasti}},
	\bibinfo{author}{\bibfnamefont{A.}~\bibnamefont{Uppstu}},
	\bibinfo{author}{\bibfnamefont{P.}~\bibnamefont{Liljeroth}},
	\bibinfo{author}{\bibfnamefont{J.}~\bibnamefont{van~der Lit}},
	\bibinfo{author}{\bibfnamefont{I.}~\bibnamefont{Swart}}, \bibnamefont{and}
	\bibinfo{author}{\bibfnamefont{A.}~\bibnamefont{Harju}},
	\bibinfo{journal}{Physical Review B} \textbf{\bibinfo{volume}{88}},
	\bibinfo{pages}{075429} (\bibinfo{year}{2013}).
	
	\bibitem[{\citenamefont{L\"{u} et~al.}(2014)\citenamefont{L\"{u}, Christensen,
			Foti, Frederiksen, Gunst, and Brandbyge}}]{Lu2014}
	\bibinfo{author}{\bibfnamefont{J.-T.} \bibnamefont{L\"{u}}},
	\bibinfo{author}{\bibfnamefont{R.~B.} \bibnamefont{Christensen}},
	\bibinfo{author}{\bibfnamefont{G.}~\bibnamefont{Foti}},
	\bibinfo{author}{\bibfnamefont{T.}~\bibnamefont{Frederiksen}},
	\bibinfo{author}{\bibfnamefont{T.}~\bibnamefont{Gunst}}, \bibnamefont{and}
	\bibinfo{author}{\bibfnamefont{M.}~\bibnamefont{Brandbyge}},
	\bibinfo{journal}{Phys. Rev. B} \textbf{\bibinfo{volume}{89}},
	\bibinfo{pages}{081405} (\bibinfo{year}{2014}).
	
	\bibitem[{\citenamefont{Soler et~al.}(2002)\citenamefont{Soler, Artacho, Gale,
			Garc\'{\i}a, Junquera, Ordej\'{o}n, and S\'{a}nchez-Portal}}]{Soler2002}
	\bibinfo{author}{\bibfnamefont{J.~M.} \bibnamefont{Soler}},
	\bibinfo{author}{\bibfnamefont{E.}~\bibnamefont{Artacho}},
	\bibinfo{author}{\bibfnamefont{J.~D.} \bibnamefont{Gale}},
	\bibinfo{author}{\bibfnamefont{A.}~\bibnamefont{Garc\'{\i}a}},
	\bibinfo{author}{\bibfnamefont{J.}~\bibnamefont{Junquera}},
	\bibinfo{author}{\bibfnamefont{P.}~\bibnamefont{Ordej\'{o}n}},
	\bibnamefont{and}
	\bibinfo{author}{\bibfnamefont{D.}~\bibnamefont{S\'{a}nchez-Portal}},
	\bibinfo{journal}{J. Phys.: Condens. Matter} \textbf{\bibinfo{volume}{14}},
	\bibinfo{pages}{2745} (\bibinfo{year}{2002}).
	
	\bibitem[{\citenamefont{Brandbyge et~al.}(2002)\citenamefont{Brandbyge, Mozos,
			Ordej\'{o}n, Taylor, and Stokbro}}]{Brandbyge2002}
	\bibinfo{author}{\bibfnamefont{M.}~\bibnamefont{Brandbyge}},
	\bibinfo{author}{\bibfnamefont{J.-L.} \bibnamefont{Mozos}},
	\bibinfo{author}{\bibfnamefont{P.}~\bibnamefont{Ordej\'{o}n}},
	\bibinfo{author}{\bibfnamefont{J.}~\bibnamefont{Taylor}}, \bibnamefont{and}
	\bibinfo{author}{\bibfnamefont{K.}~\bibnamefont{Stokbro}},
	\bibinfo{journal}{Phys. Rev. B} \textbf{\bibinfo{volume}{65}},
	\bibinfo{pages}{165401} (\bibinfo{year}{2002}).
	
	\bibitem[{\citenamefont{Perdew et~al.}(1996)\citenamefont{Perdew, Burke, and
			Ernzerhof}}]{PeBuEr.96}
	\bibinfo{author}{\bibfnamefont{J.~P.} \bibnamefont{Perdew}},
	\bibinfo{author}{\bibfnamefont{K.}~\bibnamefont{Burke}}, \bibnamefont{and}
	\bibinfo{author}{\bibfnamefont{M.}~\bibnamefont{Ernzerhof}},
	\bibinfo{journal}{Phys.~Rev.~Lett.} \textbf{\bibinfo{volume}{77}},
	\bibinfo{pages}{3865} (\bibinfo{year}{1996}).
	
	\bibitem[{\citenamefont{Paulsson et~al.}(2005)\citenamefont{Paulsson,
			Frederiksen, and Brandbyge}}]{Paulsson2005}
	\bibinfo{author}{\bibfnamefont{M.}~\bibnamefont{Paulsson}},
	\bibinfo{author}{\bibfnamefont{T.}~\bibnamefont{Frederiksen}},
	\bibnamefont{and}
	\bibinfo{author}{\bibfnamefont{M.}~\bibnamefont{Brandbyge}},
	\bibinfo{journal}{Phys. Rev. B} \textbf{\bibinfo{volume}{72}},
	\bibinfo{pages}{201101} (\bibinfo{year}{2005}).
	
	\bibitem[{\citenamefont{Frederiksen et~al.}(2007)\citenamefont{Frederiksen,
			Paulsson, Brandbyge, and Jauho}}]{Frederiksen2007}
	\bibinfo{author}{\bibfnamefont{T.}~\bibnamefont{Frederiksen}},
	\bibinfo{author}{\bibfnamefont{M.}~\bibnamefont{Paulsson}},
	\bibinfo{author}{\bibfnamefont{M.}~\bibnamefont{Brandbyge}},
	\bibnamefont{and} \bibinfo{author}{\bibfnamefont{A.-P.} \bibnamefont{Jauho}},
	\bibinfo{journal}{Phys. Rev. B} \textbf{\bibinfo{volume}{75}},
	\bibinfo{pages}{205413} (\bibinfo{year}{2007}).
	
	\bibitem[{Ine()}]{Inelastica}
	\bibinfo{howpublished}{\url{http://sourceforge.net/projects/inelastica}}.
	
	\bibitem[{\citenamefont{Sancho et~al.}(1984)\citenamefont{Sancho, Sancho, and
			Rubio}}]{Sancho1984}
	\bibinfo{author}{\bibfnamefont{M.~P.~L.} \bibnamefont{Sancho}},
	\bibinfo{author}{\bibfnamefont{J.~M.~L.} \bibnamefont{Sancho}},
	\bibnamefont{and} \bibinfo{author}{\bibfnamefont{J.}~\bibnamefont{Rubio}},
	\bibinfo{journal}{J.~Phys.~F: Met.~Phys.} \textbf{\bibinfo{volume}{14}},
	\bibinfo{pages}{1205} (\bibinfo{year}{1984}).
	
	\bibitem[{\citenamefont{Paulsson and Brandbyge}(2007)}]{Paulsson2007}
	\bibinfo{author}{\bibfnamefont{M.}~\bibnamefont{Paulsson}} \bibnamefont{and}
	\bibinfo{author}{\bibfnamefont{M.}~\bibnamefont{Brandbyge}},
	\bibinfo{journal}{Phys.~Rev.~B} \textbf{\bibinfo{volume}{76}},
	\bibinfo{pages}{115117} (\bibinfo{year}{2007}).
	
	\bibitem[{\citenamefont{Bronner et~al.}(2014)\citenamefont{Bronner, Utecht,
			Haase, Saalfrank, Klamroth, and Tegeder}}]{Bronner2014}
	\bibinfo{author}{\bibfnamefont{C.}~\bibnamefont{Bronner}},
	\bibinfo{author}{\bibfnamefont{M.}~\bibnamefont{Utecht}},
	\bibinfo{author}{\bibfnamefont{A.}~\bibnamefont{Haase}},
	\bibinfo{author}{\bibfnamefont{P.}~\bibnamefont{Saalfrank}},
	\bibinfo{author}{\bibfnamefont{T.}~\bibnamefont{Klamroth}}, \bibnamefont{and}
	\bibinfo{author}{\bibfnamefont{P.}~\bibnamefont{Tegeder}},
	\bibinfo{journal}{J.~Chem.~Phys.} \textbf{\bibinfo{volume}{140}},
	\bibinfo{pages}{024701} (\bibinfo{year}{2014}).
	
	\bibitem[{\citenamefont{Yang et~al.}(2007)\citenamefont{Yang, Park, Son, Cohen,
			and Louie}}]{Yang2007}
	\bibinfo{author}{\bibfnamefont{L.}~\bibnamefont{Yang}},
	\bibinfo{author}{\bibfnamefont{C.-H.} \bibnamefont{Park}},
	\bibinfo{author}{\bibfnamefont{Y.-W.} \bibnamefont{Son}},
	\bibinfo{author}{\bibfnamefont{M.}~\bibnamefont{Cohen}}, \bibnamefont{and}
	\bibinfo{author}{\bibfnamefont{S.}~\bibnamefont{Louie}},
	\bibinfo{journal}{Phys.~Rev.~Lett.} \textbf{\bibinfo{volume}{99}},
	\bibinfo{pages}{186801} (\bibinfo{year}{2007}).
	
	\bibitem[{\citenamefont{Jiang et~al.}(2013)\citenamefont{Jiang, Kharche, Kohl,
			Boykin, Klimeck, Luisier, Ajayan, and Nayak}}]{Jiang2013}
	\bibinfo{author}{\bibfnamefont{X.}~\bibnamefont{Jiang}},
	\bibinfo{author}{\bibfnamefont{N.}~\bibnamefont{Kharche}},
	\bibinfo{author}{\bibfnamefont{P.}~\bibnamefont{Kohl}},
	\bibinfo{author}{\bibfnamefont{T.~B.} \bibnamefont{Boykin}},
	\bibinfo{author}{\bibfnamefont{G.}~\bibnamefont{Klimeck}},
	\bibinfo{author}{\bibfnamefont{M.}~\bibnamefont{Luisier}},
	\bibinfo{author}{\bibfnamefont{P.~M.} \bibnamefont{Ajayan}},
	\bibnamefont{and} \bibinfo{author}{\bibfnamefont{S.~K.} \bibnamefont{Nayak}},
	\bibinfo{journal}{Appl. Phys. Lett.} \textbf{\bibinfo{volume}{103}},
	\bibinfo{pages}{133107} (\bibinfo{year}{2013}).
	
	\bibitem[{\citenamefont{Liang and Meunier}(2012)}]{Liang2012}
	\bibinfo{author}{\bibfnamefont{L.}~\bibnamefont{Liang}} \bibnamefont{and}
	\bibinfo{author}{\bibfnamefont{V.}~\bibnamefont{Meunier}},
	\bibinfo{journal}{Phys.~Rev.~ B} \textbf{\bibinfo{volume}{86}},
	\bibinfo{pages}{195404} (\bibinfo{year}{2012}).
	
	\bibitem[{\citenamefont{Song et~al.}(2009)\citenamefont{Song, Kim, Jang, Jeong,
			Reed, and Lee}}]{Song2009}
	\bibinfo{author}{\bibfnamefont{H.}~\bibnamefont{Song}},
	\bibinfo{author}{\bibfnamefont{Y.}~\bibnamefont{Kim}},
	\bibinfo{author}{\bibfnamefont{Y.~H.} \bibnamefont{Jang}},
	\bibinfo{author}{\bibfnamefont{H.}~\bibnamefont{Jeong}},
	\bibinfo{author}{\bibfnamefont{M.~A.} \bibnamefont{Reed}}, \bibnamefont{and}
	\bibinfo{author}{\bibfnamefont{T.}~\bibnamefont{Lee}},
	\bibinfo{journal}{Nature} \textbf{\bibinfo{volume}{462}},
	\bibinfo{pages}{1039} (\bibinfo{year}{2009}).
	
	\bibitem[{\citenamefont{Paulsson et~al.}(2008)\citenamefont{Paulsson,
			Frederiksen, Ueba, Lorente, and Brandbyge}}]{Paulsson2008}
	\bibinfo{author}{\bibfnamefont{M.}~\bibnamefont{Paulsson}},
	\bibinfo{author}{\bibfnamefont{T.}~\bibnamefont{Frederiksen}},
	\bibinfo{author}{\bibfnamefont{H.}~\bibnamefont{Ueba}},
	\bibinfo{author}{\bibfnamefont{N.}~\bibnamefont{Lorente}}, \bibnamefont{and}
	\bibinfo{author}{\bibfnamefont{M.}~\bibnamefont{Brandbyge}},
	\bibinfo{journal}{Phys.~Rev.~Lett.} \textbf{\bibinfo{volume}{100}},
	\bibinfo{pages}{226604} (\bibinfo{year}{2008}).
	
	\bibitem[{\citenamefont{Djukic and van
			Ruitenbeek}(2006)}]{RBG_Djukic_single_molecule}
	\bibinfo{author}{\bibfnamefont{D.}~\bibnamefont{Djukic}} \bibnamefont{and}
	\bibinfo{author}{\bibfnamefont{J.~M.} \bibnamefont{van Ruitenbeek}},
	\bibinfo{journal}{Nano Lett.} \textbf{\bibinfo{volume}{6}},
	\bibinfo{pages}{789} (\bibinfo{year}{2006}).
	
	\bibitem[{\citenamefont{Schneider et~al.}(2012)\citenamefont{Schneider, L\"u,
			Brandbyge, and Berndt}}]{c60}
	\bibinfo{author}{\bibfnamefont{N.~L.} \bibnamefont{Schneider}},
	\bibinfo{author}{\bibfnamefont{J.~T.} \bibnamefont{L\"u}},
	\bibinfo{author}{\bibfnamefont{M.}~\bibnamefont{Brandbyge}},
	\bibnamefont{and} \bibinfo{author}{\bibfnamefont{R.}~\bibnamefont{Berndt}},
	\bibinfo{journal}{Phys. Rev. Lett.} \textbf{\bibinfo{volume}{109}},
	\bibinfo{pages}{186601} (\bibinfo{year}{2012}).
	
	\bibitem[{\citenamefont{Engelund et~al.}(2010)\citenamefont{Engelund,
			F\"{u}rst, Jauho, and Brandbyge}}]{madse10}
	\bibinfo{author}{\bibfnamefont{M.}~\bibnamefont{Engelund}},
	\bibinfo{author}{\bibfnamefont{J.~A.} \bibnamefont{F\"{u}rst}},
	\bibinfo{author}{\bibfnamefont{A.~P.} \bibnamefont{Jauho}}, \bibnamefont{and}
	\bibinfo{author}{\bibfnamefont{M.}~\bibnamefont{Brandbyge}},
	\bibinfo{journal}{Phys. Rev. Lett.} \textbf{\bibinfo{volume}{104}},
	\bibinfo{pages}{36807} (\bibinfo{year}{2010}).
	
	\bibitem[{\citenamefont{Liu et~al.}(2012)\citenamefont{Liu, Wang, Hupalo, Lu,
			Tringides, Yao, and Ho}}]{Liu2012}
	\bibinfo{author}{\bibfnamefont{X.}~\bibnamefont{Liu}},
	\bibinfo{author}{\bibfnamefont{C.~Z.} \bibnamefont{Wang}},
	\bibinfo{author}{\bibfnamefont{M.}~\bibnamefont{Hupalo}},
	\bibinfo{author}{\bibfnamefont{W.~C.} \bibnamefont{Lu}},
	\bibinfo{author}{\bibfnamefont{M.~C.} \bibnamefont{Tringides}},
	\bibinfo{author}{\bibfnamefont{Y.~X.} \bibnamefont{Yao}}, \bibnamefont{and}
	\bibinfo{author}{\bibfnamefont{K.~M.} \bibnamefont{Ho}},
	\bibinfo{journal}{Phys. Chem. Chem. Phys.} \textbf{\bibinfo{volume}{14}},
	\bibinfo{pages}{9157} (\bibinfo{year}{2012}).
	
\end{thebibliography}
\end{document}